\documentclass{appolb}
\usepackage[utf8]{inputenc} % If utf8 encoding
\usepackage[T1]{fontenc}    %
\usepackage[english]{babel} % English, please
\usepackage[final]{microtype} % Less badboxes

\usepackage{amssymb,amsmath,mathtools} % Math
\usepackage{graphicx} % Graphics
\DeclareMathOperator*{\per}{per}  % Permanent
\DeclareMathOperator*{\Pf}{Pf}  % Pfaffian
\DeclareMathOperator{\Ai}{Ai} % Airy function
\DeclareMathOperator{\Var}{Var} % Airy function
\DeclareMathOperator{\Cov}{Cov} % Airy function
\DeclareMathOperator{\erfc}{erfc} % Complementary rror function
\DeclareMathOperator{\prob}{Prob} % Probability

\DeclarePairedDelimiter{\abs}{\lvert}{\rvert} % Absolute value
\DeclarePairedDelimiter{\average}{\langle}{\rangle} % Average
\DeclarePairedDelimiter{\inner}{\langle}{\rangle} % Inner product

\newcommand{\MeijerG}[8][\bigg]{G^{{ #2 },{ #3 }}_{{ #4 },{ #5 }} #1( \begin{matrix} #6 \\ #7 \end{matrix}\, #1\vert\, #8 #1)}
\newcommand{\hypergeometric}[6][\bigg]{\,{}_{#2} F_{#3} #1( \begin{matrix} #4 \\ #5 \end{matrix}\, #1\vert\, #6 #1)}
\newcommand{\one}{\mathbb{I}} % One?
\newcommand{\nn}{\nonumber} % No number in equations

\newcommand{\jpdf}{\text{jpdf}}
\newcommand{\meijer}{\text{Meijer}}
\newcommand{\sine}{\text{sine}}
\newcommand{\bessel}{\text{Bessel}}
\newcommand{\airy}{\text{Airy}}
\newcommand{\origin}{\text{origin}}
\newcommand{\bulk}{\text{bulk}}

\newcommand{\soft}{\text{soft}}

\newcommand{\funkybreak}%
  {\vspace*{.5\baselineskip}\centerline {$* \quad * \quad *$}\vspace*{.5\baselineskip}} % If can't make it, fake it.

\newcommand{\be}{\begin{equation}}
\newcommand{\ee}{\end{equation}}
\newcommand{\bea}{\begin{eqnarray}}  % Never use eqnarray, use align instead!
\newcommand{\eea}{\end{eqnarray}}    % eqnarray is evil!

\usepackage{hyperref}   % Internal hyperlinks
\hypersetup{
colorlinks=true,        % false: boxed links; true: colored links
linktoc=page,           % page numbers are links in the contents
linkcolor=blue,         % color of internal links (change box color with linkbordercolor)
citecolor=blue,         % color of links to bibliography
}

\begin{document}
\eqsec  % uncomment this line to get equations numbered by (sec.num)

\title{Recent exact and asymptotic results\\ for products of independent random matrices\thanks{Presented at ``Random Matrix Theory: Foundations and Applications'',\newline July 1-6, 2014, Krak\'ow, Poland}%
}
\author{Gernot Akemann and Jesper R. Ipsen
\address{Department of Physics,  Bielefeld University,\\  Postfach 100131, D-33501 Bielefeld, Germany}
\\
%{Third Author of different affiliation
%}
%the Name(s) of other Author(s)
%\address{affiliation}
}

\maketitle

\begin{abstract}
\noindent
In this review we summarise recent results for the complex eigenvalues and singular values of finite products of finite size random matrices, their correlation functions and asymptotic limits. The matrices in the product are taken from ensembles of independent real, complex, or quaternionic Ginibre matrices, or truncated unitary matrices. Additional mixing within one ensemble between matrices and their inverses is also covered. Exact determinantal and Pfaffian expressions are given in terms of the respective kernels of orthogonal polynomials or functions. Here we list all known cases and some straightforward generalisations. The asymptotic results for large matrix size include new microscopic universality classes at the origin and a generalisation of weak non-unitarity close to the unit circle. So far in all other parts of the spectrum the known standard universality classes have been identified. In the limit of infinite products the Lyapunov and stability exponents share the same normal distribution. To 
leading order they both follow a permanental point processes. Our focus is on presenting recent developments in this rapidly evolving area of research.

\end{abstract}
\PACS{02.10.Yn, 05.40.-a}   %Matrix theory, Random processes

\thispagestyle{empty}

\newpage
  
% ---------------------------------

\tableofcontents

% ---------------------------------

\raggedbottom

\section{Introduction}

\noindent
The study of products of random matrices goes back to the early days of random matrix theory, when in 1960 Furstenberg and Kesten \cite{FK60} studied them in the context of dynamical systems and their Lyapunov exponents. When multiplying matrices generically the product becomes non-Hermitian. Hence it is natural to choose each factor independently to be non-Hermitian, taken from a Gaussian distribution in the simplest case, the Ginibre ensembles \cite{Gin} with real, complex or quaternionic matrix elements. These are the analogues of the three classical Wigner--Dyson ensembles, labelled by $\beta=1,2,4$ respectively. When modelling unitary time evolution the choice of matrices to be multiplied is slightly less trivial. Taking factors from ensembles of Haar distributed orthogonal, unitary or symplectic matrices does not lead out of the one matrix case, due to the invariance of the Haar measure. 
For the product to become non-trivial instead one may choose to multiply truncated unitary matrices which are distributed according to the Jacobi measure, leading to a sub-unitary evolution. Other choices are of course possible, e.g. by multiplying unitary matrices that are not Haar distributed~\cite{JW:2004,NN:2007,BN:2008}.

In any case one has two choices in studying spectral properties of the product matrix: either the eigenvalues which are generally complex, or the singular values which are real positive. The latter were studied first in order to define and study the Lyapunov exponents. Their application to dynamical systems has lead to the activities summarised in \cite{CPV}. Examples for more recent applications include wireless communication \cite{Ralf} and combinatorics \cite{Karol}. Although in general complex eigenvalues and singular values are not related individually, in the limit of infinite products the radii and singular values become identical, in other words the stability and Lyapunov exponents have the same normal distribution \cite{ABK,Jesper2}, as first conjectured in \cite{GSO:1987}. 

It is quite surprising that the spectral properties of products of random matrices were first determined in the infinite product case, using mainly probabilistic tools, and in the limit of infinite matrix dimension (keeping the number of factors fixed) using planar Feynman diagrams \cite{BJW} and probability theory \cite{Ralf,GT,OS}, see \cite{Zrev} for a very recent review on the free probability approach.
The common point to the latter two approaches is that they contain information about global spectral properties. These were found to be universal when multiplying matrices from various different ensembles \cite{Burda}, including analytical results for factors from the elliptic ensemble \cite{ORSV}.

Apart from special cases for $2\times2$ matrices, see e.g. \cite{Andy}, explicit results for the product of $M$ matrices of size $N\times N$ with $M$ and $N$ finite are very recent, starting from \cite{ab,as,Jesper} for the complex eigenvalues and from \cite{akw,aik} for the singular values. They reveal a determinantal and Pfaffian structure and will be the main topic of this review. Inherited from the fact that the product of normal or gamma random variables is distributed according to the Meijer $G$-function \cite{Springer}, these special functions appear in the weight and determinantal expressions for the products of random matrices too.
Obviously the detailed knowledge of all eigenvalue and singular value correlation functions has opened up the possibility to study local spectral properties, and we will also review the recent progress on finding known and new universality classes. 

In \cite{ab} the joint density of complex eigenvalues for products of $M$ matrices of size $N\times N$ was derived and the kernel of orthogonal polynomials (OP) in the complex plane that determines all correlation functions was given. While in the bulk and edge large-$N$ scaling limit at fixed $M$ the respective Ginibre universality classes for a single matrix were recovered, at the origin a new universality class labelled by $M$ with a hypergeometric kernel appeared \cite{ab}. These results were generalised to products of rectangular matrices (these are also called induced Ginibre matrices \cite{Fischmann}) for $\beta=4$ in \cite{Jesper}, to the rectangular $\beta=2$ case \cite{ARRS}, and to the joint density for $\beta=1$ in \cite{IK}. Here also products of rectangular truncated unitary matrices and products of mixed type were considered, for which a weak commutation relation was proved. In all these results the change of variables from matrices to complex eigenvalues uses a generalised Schur decomposition.
 
This decomposition was well-known for $M=2$ and $\beta=1,2$, see e.g.~\cite{GL:1996}, whereas its generalisation to an arbitrary number of matrices $M$ was given in~\cite{es,ARRS} for $\beta=2$, and was extended to $\beta=4$ in~\cite{ais}. A similar extension exists for $\beta=1$ and was implicitly used in~\cite{PFreal,IK}. In \cite{ARRS} also mixed products of Ginibre and inverse Ginibre matrices, and of truncated unitary matrices and their inverses were considered, and their joint densities and corresponding kernels were given. The large-$N$ limit for such products of unitary matrices truncated form $U(N+\kappa)$ to $U(N)$, also called random contractions, was studied in \cite{ABKN}. At strong non-unitarity in the bulk and edge scaling limit the universal result for a single Ginibre matrix was recovered. In the origin limit the same new class as for products of $M$ Ginibre matrices \cite{ab} mentioned above was found. 
Very recently, a more rigorous derivation of this bulk and edge universality was presented in \cite{LW}, including the case of products of rectangular Ginibre matrices when both the inner and outer edge become soft. At weak non-unitarity the result for a single matrix \cite{KS} was extended to a new kernel labelled by $M\kappa$ in \cite{ABKN}. A discussion for $M>1$ products of truncated orthogonal (and unitary symplectic) matrices was given in~\cite{IK}, albeit an understanding of the universal kernels is still lacking, for a single truncated orthogonal matrix see however \cite{KS2}. 

The distribution of radii of complex eigenvalues of products of Ginibre matrices is given by permanents \cite{as,ais}, cf. \cite{Hough} for $M=1$. The corresponding gap and overcrowding probabilities were derived in \cite{as} for $\beta=2$ and in \cite{ais} for $\beta=4$, including the corresponding asymptotic expansions for finite and infinite point processes, for fixed $M$. The limiting distribution of the largest radius of all eigenvalues was shown to interpolate between the log-normal and Gumbel distribution (valid for a single Ginibre matrix \cite{Rider}) in a double scaling limit in \cite{JQ}, where both $N$ and $M$ become infinite.

% -----------------------------------

Turning to singular values their joint density and corresponding kernel of orthogonal functions were derived in \cite{akw} for square and \cite{aik} for rectangular matrices with $\beta=2$. These enjoy 
a relation to multiple OP \cite{lun} and a universal correlation kernel labelled by $M$ was found in the local origin scaling limit \cite{AZ}. For $M=2$ it agrees with the limiting kernel in a Cauchy two-matrix model \cite{MB1}, a correspondence that was extended to the Cauchy multi-matrix model and general $M$ very recently \cite{MB2}, cf. \cite{FK:2014}.
This result was generalised in \cite{ArnoDries} to a mixed product of Ginibre matrices times a single truncated unitary matrix, introducing the notion of polynomial ensembles. There, their relation to a certain type of biorthogonal ensembles previously studied in~\cite{Muttalib:1995,Borodin} was pointed out. Polynomial ensembles enjoy special invariance properties, related to products of random matrices \cite{Kuijlaars:2015}.
The same limiting kernel as in \cite{AZ} was also found in \cite{PF14} for products of Ginibre and inverse Ginibre matrices, where the average characteristic polynomial was computed as well. 
The zeros and asymptotic analysis of the average characteristic polynomial was performed in \cite{TD}.
The Fredholm determinant for the gap probability at the origin was shown to satisfy a system of non-linear ordinary differential equations \cite{ES14}.
In the bulk and at the soft edge it was conjectured in \cite{akw} to find the universal sine- and Airy kernel respectively, see \cite{lundang} for a very recent proof including mixed products of Ginibre matrices and their inverse.
The correlation functions of singular values of truncated unitary matrices is also currently under way \cite{KKS}.
So far information on local properties of the singular values for the $\beta=1,4$ classes has been very difficult to access, due to absence of the corresponding Harish-Chandra integral. Very recent progress has been possible using supersymmetric techniques, see \cite{Mario15}.

The limiting positions for the Lyapunov exponents of products of $\beta=2$ \cite{PF13} and $\beta=4$ \cite{Kargin} Ginibre matrices were derived using probabilistic methods, including correlated Gaussian distributions for each factor. The detailed knowledge for the joint densities described above allowed to show \cite{ABKN} for uncorrelated complex Ginibre matrices, that each Lyapunov exponent becomes normally distributed, following a permanental point process. The same leading order behaviour holds taking the identical limit of the radii of complex eigenvalues, called stability exponents. This leads to a one-to-one correspondence between limiting radii and singular values.   
The expressions for the stability exponents were extended to $\beta=4$ and $\beta=1$ most recently in \cite{Jesper2}, where the latter case was obtained under the assumption that for large $M$ all eigenvalues of the real product matrix become real. Such a behaviour was previously observed in \cite{arul} and proved for general $N$ in \cite{PFreal}.

All aforementioned finite-$N$ and -$M$ results have assumed that each factor in the product is independent\footnote{Of course powers of a single matrix form an exception, leading to the same global density as the product of independent matrices, see e.g. \cite{GT}.}. To date only complex eigenvalues of products of $M=2$ matrices that are coupled through an Itzykson--Zuber term and thus not independent have been studied for finite-$N$ at $\beta=2$ \cite{James}, $\beta=4$ \cite{a05}, and $\beta=1$ \cite{aps}. They are given in terms of parameter dependent families of Laguerre polynomials in the complex plane, and we refer to the review \cite{A05acta} where all these results are summarised. Results for singular values of products of such dependent matrices are currently under way \cite{as2}.

The remaining content of this paper is organised as follows. In section \ref{finiteNM} we will present exact results for finite $N$ and $M$ including the joint densities, kernels and correlation functions; the section is divided in to two subsections \ref{EV} and \ref{SVfinite} discussing results for complex eigenvalues and singular values, respectively. In both cases, we consider Ginibre and inverse Ginibre matrices as well as truncated unitary matrices. We turn to the local large-$N$ limits at fixed $M$ depending on the location in the spectrum in section~\ref{largeN}. Here we will be very brief and only give more details for the limiting kernels that were not previously known. Again, we will first discuss complex eigenvalues in subsection~\ref{EVlocal} and then discuss singular values in subsection~\ref{SVasymp}. 
The following section \ref{largeM} is devoted to the opposite infinite product limit at fixed $N$. 
Our discussion of open problems follows in section \ref{concs}.

\section{Exact results for finite products of finite size matrices}
\label{finiteNM}

\subsection{Complex eigenvalues}
\label{EV}

% -------------------------------------------------

\subsubsection{Ginibre matrices}
\label{EVginibre}

\noindent
We begin with the simplest case: A \emph{product of rectangular Ginibre matrices}. We emphasise that this special case is very illustrative, since the main ideas from the treatment of products of Ginibre matrices extend to all the other examples described in this review.  

We are seeking the statistical properties of complex eigenvalues of the product matrix
\begin{equation}
\Pi_M\equiv X_M X_{M-1}\cdots X_2 X_1
\label{prod}
\end{equation}
where each $X_j$ is an $N_j\times N_{j-1}$ matrix distributed according to a Gaussian density
\begin{equation}
P(X_j)=\left(\frac{\beta}{2\pi}\right)^{\beta N_j N_{j-1}/2}\exp\left[-\frac{\beta}{2\gamma}\Tr X_j X_j^\dag\right], \ j=1,\ldots,M.
\label{Ginibre}
\end{equation}
Here the matrices $X_j$ ($j=1,\ldots,M$) are independent and the index $\beta=1,2,4$ denotes whether real, complex or quaternionic matrices are considered (we only multiply matrices with the same $\beta$). The parameter
\begin{equation}
\gamma=
\begin{cases}
1 & \text{for}\quad\beta=1,2 \\
2 & \text{for}\quad\beta=4
\end{cases}
\label{gammadef}
\end{equation}
is related to the fact that quaternions are given by their $2\times 2$ matrix representation, which implies that the eigenvalues come in complex conjugate pairs (see e.g.~\cite{Mehta}).
If we disregard eigenvalues which are trivially zero, then we can choose $N\equiv N_0\leq N_1\leq \cdots\leq N_M$ without loss of generality; this is due to a weak commutation relation~\cite{IK}. Furthermore, we can parametrise the matrix product as
\begin{equation}
X_M X_{M-1}\cdots X_2 X_1=U\begin{pmatrix} \tilde{X}_M \tilde{X}_{M-1}\cdots \tilde{X}_2 \tilde{X}_1 \\ 0 \end{pmatrix},
\end{equation}
where $U$ is an $N_M\times N_M$ orthogonal ($\beta=1$), unitary ($\beta=2$), or unitary symplectic ($\beta=4$) matrix and each $\tilde X_j$ is an $N\times N$ matrix. This parameterisation results in the measure~\cite{IK}
\begin{equation}
\prod_{j=1}^M dX_j P(X_j)=d\mu(U)\prod_{j=1}^M d\tilde{X}_j\tilde P_{\nu_j}(\tilde{X}_j)\ ,
\label{squareReduction}
\end{equation}
where $dX_j$ and $d\tilde X_j$ are the flat measures over all independent matrix elements, $d\mu(U)$ is the normalised Haar measure, and each $\tilde X_j$ is distributed according to the induced density
\begin{equation}
\tilde P_{\nu_j}(\tilde{X}_j)\propto\det[\tilde{X}_j \tilde{X}_j^\dag]^{\beta\nu_j/(2\gamma)}\exp\Big[-\frac{\beta}{2\gamma}\Tr \tilde{X}_j \tilde{X}_j^\dag\Big].
\label{induced}
\end{equation}
Here we have introduced the convenient notation $\nu_j\equiv N_j-N$ for the differences between matrix dimensions. We emphasise that~\eqref{squareReduction} enables us to reduce the problem involving rectangular matrices to the problem involving square matrices, which is a considerable simplification (we refer to~\cite{IK} for a more thorough discussion). It should be noted that this simplification is possible for all examples mentioned in this review. Finally, we note that the induced densities~\eqref{induced} are isotropic, i.e. they are invariant under $X_j\to UX_jV$, where $U$ and $V$ are orthogonal ($\beta=1$), unitary ($\beta=2$), or unitary symplectic ($\beta=4$) matrices. This is an extremely important observation needed for the weak communication relation~\cite{IK}, which states that any averaged property of such a product matrix (depending only on the product matrix itself, not on the individual matrices) is independent of the ordering of the factors. For products of rectangular Ginibre matrices 
this implies that 
all averaged properties (e.g. correlations between eigenvalues or singular values) are invariant under permutations of the indices $\nu_m$ ($m=1,\ldots,M$). The reader might note that all products we consider in this review are independent of the ordering of the factors. This originates from the fact that \emph{all ensembles we consider are isotropic}. 

\funkybreak

\noindent
We are now ready to state the normalised \emph{joint probability density function} (jpdf) of complex eigenvalues for the product matrix $\Pi_M$. For $\beta=2$ the jpdf reads~\cite{ab,ARRS,IK}
\begin{equation} 
{\cal P}_{\jpdf,\nu}^{\beta=2}(z_1,\ldots,z_N)= \frac{1}{{\cal Z}_{N,\nu}^{\beta=2}} \prod_{n=1}^N w_\nu^{\beta=2}(z_n) \prod_{1\leq j<l\leq N}|z_l-z_j|^2.
\label{jpdf2}
\end{equation}
Apart from the corresponding \emph{weight function} given by a Meijer $G$-function
\begin{equation} 
w_\nu^{\beta}(z)=
\MeijerG{M}{0}{0}{M}{-}{\frac{\beta}{2}\nu_1, \ldots, \frac{\beta}{2}\nu_M}{\left(\frac{\beta}{2}\right)^M\abs z^2},
\label{wGdef}
\end{equation}
it agrees with the jpdf for a single Ginibre matrix $M=1$, where the repulsion of eigenvalues is given by the squared absolute value of the Vandermonde determinant.
The \emph{normalisation constant} is given for all three $\beta=1,2,4$ by 
\begin{equation} 
{\cal Z}_{N,\nu}^{\beta}=\frac{N!\pi^{N(\beta-\gamma)/\gamma}}{2^{(2-\beta)\gamma MN(N+1)/4}}\prod_{n=1}^N\prod_{j=1}^M\Gamma\left(\frac{\beta}{2}(\nu_j+n)\right).
\label{Zdef}
\end{equation}
The following steps are needed to derive the jpdf in~\eqref{jpdf2} for $\beta=2$ for general $M$ (and subsequently for the other $\beta$'s): i) a generalised Schur decomposition \cite{ab,es,ARRS} of the individual factors distributed according to~\eqref{induced}, and ii) the fact that the upper triangular matrices from this decomposition decouple. The remaining integrals can be performed \cite{ab,ARRS,IK} and lead to the Meijer $G$-function as the weight in~\eqref{wGdef}.
For the most general set of indices the Meijer $G$-function is defined as 
\cite{Bateman:1953}
\begin{equation}
\MeijerG{m}{n}{p}{q}{a_1,\ldots, a_p}{b_1, \ldots,  b_q}{z}
\equiv\frac{1}{2\pi i}\int_{\cal C}\frac{ds\,z^s\prod_{j=1}^m\Gamma(b_j-s)\prod_{j=1}^n\Gamma(1-a_j+s)}{\prod_{j=m+1}^q\Gamma(1-b_j+s)\prod_{j=n+1}^p\Gamma(a_j-s)}.
\label{Gdef}
\end{equation}
%%%%%%% contour specified
The integration contour ${\mathcal C}$ depends on the poles of the Gamma functions, cf. \cite{Bateman:1953}. In the simplest case with $p+q<2(m+n)$ and $z$ in the upper half plane it runs from $-i\infty$ to $+i\infty$ leaving all poles of $\Gamma(b_j-s)$ to the right and of $\Gamma(1-a_j+s)$ to the left. Empty products are defined as unity.
For the special case~\eqref{wGdef} an alternative multiple integral representation of the Meijer $G$-function exists~\cite{ab,ARRS}
\begin{equation}
\MeijerG{M}{0}{0}{M}{-}{\nu_1, \ldots, \nu_M}{\abs z^2}=\pi^{1-M}\prod_{j=1}^M\int_{\mathbb C}d^2z_j|z_j|^{2\nu_j}e^{-|z_j|^2}  
\delta^{2}(z-z_M\cdots z_1).
\label{int1}
\end{equation}
Note that the right hand side corresponds to the density of a product of (unnormalised) scalar-valued random variables. Here we get the explanation why the Meijer $G$-function appears: it was previously known for products of Gaussian or gamma random variables that their products is distributed according to the Meijer $G$-function \cite{Springer}. This property is passed on to products of random matrices as well.
It follows immediately from the representation~\eqref{int1} that the (bi-)moments of the weight functions are given by
\begin{equation}
\int_{\mathbb C}d^2z\, w_\nu^{\beta=2}(z)\,z^{k}z^{*\,\ell}=\delta_{k,\ell}\,\pi \prod_{j=1}^M\Gamma(\nu_j+k+1).
\label{moments}
\end{equation}
Here the moments are zero for $k\neq \ell$ since the weight function~\eqref{wGdef} is invariant under rotation in the complex plane.

\funkybreak

\noindent
The \emph{$k$-point density correlation functions} are defined following \cite{Mehta}, 
\begin{equation}
R_k^{\beta}(z_1,\ldots,z_k)\equiv \frac{N!}{(N-k)!}
\int_{\mathbb C} d^2z_{k+1}\ldots   d^2z_{N} {\cal P}^{\beta}_{\jpdf}(z_1,\ldots,z_N).
\label{Rkdef}
\end{equation}
We will consider jpdfs with the following form:
\begin{equation} 
\mathcal P_{\jpdf}^{\beta=2}(z_1,\ldots,z_N)= \frac{1}{\mathcal Z_{N}^{\beta=2}} \prod_{n=1}^N w^{\beta=2}(\abs{z_n}) \prod_{1\leq \ell<k\leq N}|z_k-z_\ell|^2,
\label{jpdf2general}
\end{equation}
where $w^{\beta=2}(\abs z)$ is an arbitrary weight function that only depends on the modulus. Then it is well-known that the corresponding OPs are monomials,
\begin{equation}
\int_{\mathbb C} d^2z\ w^{\beta=2}(|z|) z^k z^{*\,\ell}=\delta_{k,\ell}h_k.
\label{OPdef}
\end{equation}
Furthermore, it follows from general considerations that the correlation functions including the jpdf for $k=N$ form a \emph{determinantal point process} \cite{Mehta}
\begin{equation}
R_k^{\beta=2}(z_1,\ldots,z_k)
=\prod_{j=1}^k w^{\beta=2}(|z_j|)
\det_{1\leq i,j,\leq k}\left[K_N^{\beta=2}(z_i,z_j)\right],
\label{Rkb=2}
\end{equation}
where the \emph{kernel} is given by
\begin{equation}
K_N^{\beta=2}(z,u)=\sum_{n=0}^{N-1} \frac{(zu^*)^n}{h_n}.
\label{KNdef}
\end{equation}
Here $h_n$ denotes the squared norm determined through the orthogonality relation~\eqref{OPdef}. The normalisation constant in~\eqref{jpdf2general} is uniquely determined by these squared norms as well. Explicitly, we have
\begin{equation}
\mathcal Z_N^\beta=N!\prod_{n=0}^{N-1}h_n.
\label{ZNgeneral}
\end{equation}
For the product of complex ($\beta=2$) rectangular Ginibre matrices the jpdf is given by~\eqref{jpdf2} with weight~\eqref{wGdef} and the squared norms are immediately obtained from~\eqref{moments}. Thus the \emph{kernel for the product of rectangular Ginibre matrices} is given by~\cite{ab,ARRS}
\begin{equation} 
K_{N,\nu}^{\beta=2}(z,u)=\frac1\pi 
\sum_{n=0}^{N-1} \frac{(zu^*)^n}{\prod_{m=1}^M\Gamma(n+1+\nu_m)}\ ,
\label{Kgauss2}
\end{equation}
while the corresponding \emph{normalisation constant} is
\begin{equation}
\mathcal Z_{N,\nu}^\beta=N!\,\pi^M\prod_{j=1}^M\Gamma(N+1+\nu_j)\ .
\end{equation}
This determines all $k$-point correlation functions for $\beta=2$ via~(\ref{Rkb=2}) with weight (\ref{wGdef}). 

The second set of results that follows from eq. (\ref{jpdf2general}) for $\beta=2$ is the \emph{joint density of the radii} $r_j$ of the complex eigenvalues, $z_j=r_je^{i\theta_j}$. It is obtained by  integrating over all angles $\theta_j$ leading to a permanent~\cite{as}:
\begin{equation} 
\prod_{n=1}^N\int_{0}^{2\pi}d\theta_nr_n
{\mathcal P}_{\jpdf}^{\beta=2}(z_1,\ldots,z_N)= 
\frac{\prod_{n=1}^N 2\pi\, w^{\beta=2}(r_n)}{{\cal Z}_{N}^{\beta=2}}
\per_{1\leq j,l\leq N}\left[r_{j-1}^{2l-1}\right],
\label{jpdf2rad}
\end{equation}
where we have included the factors $r_n$ from the radial measure $dr_nr_n$. 
The radii thus become independent random variables, generalising the result in \cite{Kostlan}. Moreover, the \emph{hole probability} that a disc of radius $r$ centred at the origin is empty of eigenvalues becomes
\begin{align}
\prob[\forall j:  r_j>r]
&\equiv\prod_{n=1}^N\int_{r}^\infty dr_{n}r_n\int_0^{2\pi}d\theta_{n} {\cal P}^{\beta=2}_{\jpdf}(z_1,\ldots,z_N)
\label{holedef}  \\
&=\prod_{n=1}^N \frac{\MeijerG{M+1}{0}{1}{M+1}{1}{0,n+\nu_1, \ldots, n+\nu_M}{r^2}}{\prod_{j=1}^M\Gamma(n+\nu_j)}\ .
\label{holeb2}
\end{align}
The second line was derived in \cite{ais}. It agrees with \cite{Mehta} for $M=1$ with $\nu_1=0$. An alternative result for $M=2$ was previously obtained in~\cite{apshi}.

\funkybreak

\noindent
The quaternionic \emph{$k$-point correlation functions} for $\beta=4$ are defined identically to the complex case~\eqref{Rkdef}. Here we will consider jpdfs of the form:
\begin{equation} 
{\mathcal P}_{\jpdf}^{\beta=4}(z_1,\ldots,z_N)= \frac{1}{{\cal Z}_{N}^{\beta=4}} \prod_{n=1}^N w^{\beta=4}(\abs{z_n})|z_n-z_n^*|^2 \prod_{1\leq k<\ell\leq N}|z_\ell-z_k|^2|z_\ell-z_k^*|^2,
\label{jpdf4general}
\end{equation}
where the weight function is invariant under rotation in the complex plane. We restrict ourselves to the upper half plane ${\mathbb C}_+$, due to complex conjugated pairing of eigenvalues for $\beta=4$. It should be noted that except for the weight function the jpdf is identical to that of a single Ginibre matrix $M=1$. Additionally, we define the \emph{skew-symmetric product}
\begin{equation}
\inner{f, g}_s\equiv \int_{\mathbb C_+}d^2z\,w^{\beta=4}(\abs{z})(z^*-z)
\big(f(z)g(z^*)-f(z^*)g(z)\big).
\label{skewProduct}
\end{equation}
It follows from general considerations that the corresponding $k$-point correlation functions including the jpdf form a \emph{Pfaffian point process}~\cite{Mehta,Kanzieper:2002}
\begin{multline}
R_k^{\beta=4}(z_1,\ldots,z_k)
=\\
\prod_{j=1}^k w^{\beta=4}(z_j)(z_j^*-z_j)
\Pf_{1\leq i,j\leq k}
\begin{bmatrix}
K_{N}^{\beta=4}(z_i,z_j^*) & -K_{N}^{\beta=4}(z_i^*,z_j^*)\\
K_{N}^{\beta=4}(z_i,z_j) & -K_{N}^{\beta=4}(z_i^*,z_j)
\end{bmatrix} \ ,
\label{Rkb=4}
\end{multline}
with the \emph{kernel} given in terms of monic skew-orthogonal polynomials
\begin{equation}
K_{N,\nu}^{\beta=4}(z,u) =\sum_{n=0}^{N-1}\frac{p_{2n+1}(z)p_{2n}(u)-p_{2n+1}(v)p_{2n}(z)}{h_n}.
\label{Knb=4}                                                      
\end{equation}
Here the \emph{skew-orthogonal polynomials} $p_n(z)$ are defined such that they satisfy the \emph{skew-orthogonality relations}~\cite{Mehta,Kanzieper:2002}
\begin{equation}
\inner{p_{2k+1},p_{2\ell+1}}_s=\inner{p_{2k},p_{2\ell}}_s=0,\qquad
\inner{p_{2k+1},p_{2\ell}}_s=h_k\delta_{k,\ell}.
\end{equation}
We know that the weight function in the jpdf~\eqref{jpdf4general} is invariant under rotation in the complex plane. This implies that
\begin{equation}
\int_{\mathbb C_+} d^2z\ w^{\beta=4}(|z|) z^k z^{*\,\ell}=s_k\,\delta_{k,\ell},
\label{OP4}
\end{equation}
where $s_k$ are constants depending on the weight. It was pointed out in~\cite{Jesper} that in order to find the skew-orthogonal polynomials it is sufficient to calculate the constants, $s_k$. Explicitly, we have
\begin{equation}
p_{2n}(z)=\sum_{k=0}^n \bigg[\prod_{\ell=k+1}^n\frac{s_{2\ell}}{s_{2\ell-1}}\bigg]z^{2k},\ \ 
p_{2n+1}(z)=z^{2n+1},\ \ 
h_{n}=2\,s_{2n+1}.
\label{skewOPgeneral}
\end{equation}
The main idea behind this result is 
that the odd polynomials are always monomials, 
due to the rotational invariance of the weight.

We can now turn to \emph{products of rectangular Ginibre matrices with quaternionic matrix elements} ($\beta=4$), where we refer to \cite{Jesper,IK,ais} for details. 
The derivation of the jpdf for the complex eigenvalues is based on the generalised Schur decomposition for quaternionic matrices, see~\cite{ais}. With the notation given above, we have
\begin{equation} 
{\mathcal P}_{\jpdf,\nu}^{\beta=4}(z_1,\ldots,z_N)= \frac{1}{{\cal Z}_{N,\nu}^{\beta=4}} \prod_{n=1}^N w_\nu^{\beta=4}(z_n)|z_n-z_n^*|^2 \prod_{1\leq k<l\leq N}|z_l-z_k|^2|z_k-z_l^*|^2,
\label{jpdf4}
\end{equation}
with weight and normalisation constant defined in (\ref{wGdef}) and (\ref{Zdef}), respectively. Inserting~\eqref{wGdef} with $\beta=4$ into~\eqref{OP4} yields
\begin{equation}
s_{n}=\frac{\pi}{2^{M(n+1)+1}}\prod_{m=1}^M\Gamma(2\nu_m+n+1).
\end{equation}
This determines the skew-orthogonal polynomials~\eqref{skewOPgeneral} and therefore all correlation functions via~\eqref{Rkb=4}, with weight and kernel given by~\eqref{wGdef} and~\eqref{Knb=4}, respectively.

Turning to the \emph{distribution of radii} we obtain up to factors of 2 the same result as in eq. (\ref{jpdf2rad}) after integration 
\begin{equation} 
\prod_{n=1}^N\int_{0}^{2\pi}d\theta_n r_n
{\cal P}_{\jpdf,\nu}^{\beta=4}(z_1,\ldots,z_N)= 
\frac{
\prod_{n=1}^N 4\pi\, w_\nu^{\beta=4}(r_n)}{{\cal Z}_{N,\nu}^{\beta=4}}
\per_{1\leq j,l\leq N}\left[r_{j-1}^{4l-3}\right]\ .
\label{jpdf4rad}
\end{equation}
This is true despite the initial repulsion of eigenvalues from the real line in eq. (\ref{jpdf4}). The radii are once again independent random variables, generalising the results of \cite{Rider}. 
Likewise the $\beta=4$ \emph{hole probability} defined as in eq. (\ref{holedef}) was obtained in \cite{ais},
\begin{equation}
\prob[\forall j:  r_j>r]=\prod_{n=1}^N\frac{\MeijerG{M+1}{0}{1}{M+1}{1}{0,2n+2\nu_1, \ldots, 2n+2\nu_M}{2^Mr^2}}{\prod_{j=1}^M\Gamma(2n+2\nu_j)} .
\label{holeb4}
\end{equation}
For $M=1$ it agrees with \cite{Mehta}; for $M=2$
see~\cite{apshi} for a different expression. 

\funkybreak

\noindent
Finally, we state the jpdf for \emph{products of rectangular real Ginibre matrices} ($\beta=1$). It is well-known that the eigenvalues of a real matrix are either real or come in complex conjugate pairs. The main difficulty for real matrices is that a complete triangularisation is possible if and only if all the eigenvalues are real~\cite{GL:1996}. Typically, a real Ginibre matrix will have both real and complex eigenvalues which prevents this. However, it is always possible to make an incomplete triangularisation involving $2\times 2$ matrices. If $N$ is even then it is possible to write down the \emph{jpdf in terms of $2\times 2$ matrices}~\cite{IK},
\begin{equation} 
{\cal P}_{\jpdf,\nu}^{\beta=1}(Z_1,\ldots,Z_{N/2})= 
\frac{1}{{\cal Z}_{N,\nu}^{\beta=1}} \prod_{n=1}^{N/2} W_\nu^{\beta=1}(Z_n) \prod_{1\leq k<\ell\leq \frac N2}\abs[\big]{\det[Z_\ell\otimes\one_2-\one_2\otimes Z_k]},
\label{jpdf1}
\end{equation}
where each $Z_n$ is a real $2\times 2$ matrix, hence its eigenvalues are either real or a complex conjugate pair. These eigenvalues can be identified with the eigenvalues of the original $N\times N$ matrix $\Pi_M$. A similar expression for $N$ odd was given in~\cite{IK}, but will not be repeated here. The \emph{weight function with a matrix argument} is given by
\begin{equation}
W_\nu^{\beta=1}(Z)=\prod_{j=1}^M\int_{\mathbb R^{2\times 2}}dZ_j|\det Z_j|^{\nu_j}e^{-\frac12\Tr Z_j^TZ_j}
\delta^{2\times 2}(Z-Z_M\cdots Z_1),
\label{w22b=1}
\end{equation}
which should be compared to the expression~\eqref{int1}. The fact that the jpdf~\eqref{jpdf1} is only known up to $2\times2$ matrix integrals plus the fact that one has to distinguish real and complex conjugate eigenvalue pairs has so far prevented progress in computing density correlation functions or the distribution of radii. In~\cite{IK}, a general approach was presented using that the eigenvalues of a $2\times 2$ matrix can be linked to its singular values. This approach led to an $M$-fold integral representation of the two-point weight, but the expression was too complicated for any further calculations to be tractable.

As mentioned above, a complete triangularisation is possible when all eigenvalues are real, $z_j=x_j\in\mathbb{R}$ ($j=1,\ldots,N$). In this special case, the approach used for $\beta=2,4$ can be extended in a straightforward manner leading to a \emph{jpdf with all eigenvalues real}~\cite{PFreal},
\begin{equation} 
{\cal P}_{\jpdf,\nu}^{\beta=1,\mathbb R}(x_1,\ldots,x_N)= \frac{1}{{\cal Z}_{N,\nu}^{\beta=1}} \prod_{n=1}^N w_\nu^{\beta=1}(x_n) \prod_{1\leq j<l\leq N}|x_l-x_j|,
\label{jpdf1real}
\end{equation}
with the weight and normalisation given in eqs. (\ref{wGdef}) and (\ref{Zdef}), respectively. If $N$ is even then the probability that all eigenvalues are real is given by~\cite{PFreal}
\begin{multline}
\prod_{n=1}^N\int_{\mathbb R}dx_n{\cal P}_{\jpdf,\nu}^{\beta=1,\mathbb R}(x_1,\ldots,x_N)= \\
\frac{\det\limits_{1\leq k<\ell\leq\frac N2}\left[\MeijerG{M+1}{M}{M+1}{M+1}{\frac32-\frac{\nu_1}2-\ell,\ldots,\frac32-\frac{\nu_M}2-\ell,1}{0,\frac{\nu_1}2+k,\frac{\nu_M}2+k}{1}\right]}
{\prod_{n=1}^N\prod_{m=1}^M\Gamma(\frac{\nu_m}2+\frac n2)}.
\end{multline}
A similar expression holds for $N$ odd.

% -----------------------------------------------

\subsubsection{Ginibre and inverse Ginibre matrices}
\label{EVinv}

\noindent
Next we consider \emph{the generalised eigenvalue problem} given by the characteristic equation
\begin{equation}
\det[Y_{1}\cdots Y_L\lambda-X_{M}\cdots X_1]=0\ ,
\label{charEq}
\end{equation}
where each $X_j$ is an $N_j\times N_{j-1}$ matrix and each $Y_j$ is an $N_{M+j-1}\times N_{M+j}$ matrix. If each $X_j$ and $Y_j$ is distributed independently according to~\eqref{Ginibre}, then the generalised eigenvalue problem may formally be thought of finding the eigenvalues of a \emph{mixed product of Ginibre and inverse Ginibre matrices},
\begin{equation}
(Y_1\cdots Y_L)^{-1}X_M\cdots X_1.
\end{equation}
This was studied for square matrices in~\cite{ARRS}; the \emph{jpdf of complex eigenvalues} reads
\begin{equation} 
{\mathcal P}_{\jpdf,\nu,\mu}^{\beta=2}(z_1,\ldots,z_N)= \frac{1}{{\cal Z}_{N,\nu,\mu}^{\beta=2}} \prod_{n=1}^N w_{\nu,\mu}^{\beta=2}(z_n) \prod_{1\leq j<l\leq N}|z_l-z_j|^2,
\label{jpdf2GG}
\end{equation}
with \emph{weight function} 
\begin{equation} 
w_{\nu,\mu}^{\beta=2}(z)=\MeijerG{M}{L}{L}{M}{-N-\mu_1,\ldots,-N-\mu_L}{\nu_1,\ldots,\nu_M}{\abs z^2}.
\label{wGG}
\end{equation}
Here we use the notation $\nu_i=N_i-N$ for $i=1,\ldots,M$ and $\mu_j=N_{M+j}-N$ for $j=1,\ldots,L$. Equivalently the weight function can be written in the following integral representation
\begin{align}
w_{\nu,\mu}^{\beta=2}(z)=\pi^{1-M-L}
\prod_{\ell=1}^L&\int_{\mathbb C} d^2\,u_\ell e^{-\abs{u_\ell}^2}\abs{u_\ell}^{2(N+\mu_\ell-2)} \nn\\
\times \prod_{m=1}^M&\int_{\mathbb C} d^2z_m\, e^{-\abs{z_m}^2}\abs{z_m}^{2\nu_m}\,
\delta^2\left(z-\frac{z_M\cdots z_1}{u_L\cdots u_1}\right).
\end{align}
It follows that the squared norms defined via~\eqref{OPdef} are given by
\begin{equation}
h_n^{\nu,\mu}=\pi\prod_{m=1}^M\Gamma(\nu_m+n+1)\prod_{\ell=1}^L\Gamma(N+\mu_\ell-n).
\label{moments2}
\end{equation}
Thus the \emph{normalisation constant} directly follows from~\eqref{ZNgeneral}, while all $k$-point correlation functions are given via~(\ref{Rkb=2}) with weight~(\ref{wGG}) and kernel~\eqref{KNdef}.
Note that in both~\eqref{ZNgeneral} and~\eqref{KNdef} we have $0\leq n\leq N-1$, which ensures that the gamma functions in~\eqref{moments2} are well-defined.
For $L=0$, the jpdf~\eqref{jpdf2GG} reduces to the previous results~(\ref{jpdf2}), while for $M=L=1$ it reduces to the spherical ensemble introduced and solved in \cite{Krishnapur}. 

The joint density of the radii is given by~(\ref{jpdf2rad}) and a simple calculation starting from the definition~(\ref{holedef}) yields the \emph{hole probability}
\begin{equation}
\prob[\forall j: r_j>r]=
\prod_{n=1}^N\frac{\MeijerG{M+1}{L}{L+1}{M+1}{n-N-\mu_1,\ldots,n-N-\mu_L,1}{0,n+\nu_1, \ldots, n+\nu_M}{r^2}}{\prod_{k=1}^M\Gamma(n+\nu_k+1)\prod_{\ell=1}^L\Gamma(N+\mu_\ell-n)}.
 \label{holeb2GG}
\end{equation}

\funkybreak

\noindent
The result for $\beta=2$ can be extended to $\beta=4$ in a simple manner. The \emph{jpdf of complex eigenvalues} is of the same form as~\eqref{jpdf4general} except that the weight function and the normalisation constant changes. The \emph{weight function} for $\beta=4$ can be expressed in terms of~\eqref{wGG},
\begin{equation}
w_{\nu,\mu}^{\beta=4}(z)=w_{2\nu,2\mu+N}^{\beta=2}(2^{(M-L)/2}z).
\end{equation}
It follows from~\eqref{OP4} that
\begin{equation}
s_{n}=\frac{\pi}{2^{(M-L)(n+1)+1}}\prod_{m=1}^M\Gamma(2\nu_m+n+1)\prod_{\ell=1}^L\Gamma(2N+2\mu_m-n+1).
\end{equation}
This determines the skew-orthogonal polynomials~\eqref{skewOPgeneral} and therefore all correlation functions via~\eqref{Rkb=4} with the kernel given by~\eqref{Knb=4}. The corresponding spherical ensemble ($M=L=1$) for $\beta=4$ were studied in \cite{Anthony}. For $\beta=1$, the spherical ensemble was studied in~\cite{FM} but we will not discuss the generalisation to arbitrary $M$ and $L$ for $\beta=1$ here.                                                             

% --------------------------------------------------------

\subsubsection{Truncated unitary matrices}
\label{EVtrunc}

\noindent
Now we turn to \emph{products of truncated unitary matrices}. Consider $M$ independent orthogonal ($\beta=1$), unitary ($\beta=2$), or unitary symplectic ($\beta=4$) matrices $U_j$ of different sizes $K_j$ for $j=1,\ldots,M$. Let the unitary matrices be uniformly distributed with respect their corresponding Haar measure; we seek the truncation of each matrix $U_j$ to its upper-left sub-block $X_j$ of size $N_j\times N_{j-1}$. We are interested in the complex eigenvalues of the following product of $M$ truncated unitary matrices
\begin{equation} 
\Pi_M\equiv X_M\cdots X_1.
\label{prodUU}
\end{equation}
For $K_j-N_j-N_{j-1}>0$, projection of the Haar measure on the original group to a measure for the sub-block gives
\begin{equation}
P_j(X_j)\propto \Theta[1-X_j^\dagger X_j]\det[1-X_j^\dagger X_j]^{\beta(K_j-N_j-N_{j-1}+1-2/\beta)/(2\gamma)}\ ,
\label{TruncUniDens}
\end{equation}
for each $X_j$. Here $\Theta$ denotes the Heaviside theta function of matrix argument. In the more general case see \cite{ABKN} for an integral representation of the measure.

For $\beta=2$, the \emph{jpdf of complex eigenvalues} derived in \cite{ARRS,IK,ABKN} is of the same form as eq. (\ref{jpdf2}), 
with the \emph{weight function} given by
\begin{equation} 
w_{\nu,\kappa}^{\beta=2}(z)=\MeijerG{M}{0}{M}{M}{\kappa_1,\ldots,\kappa_M}{\nu_1,\ldots,\nu_M}{\abs z^2}.
\label{wU}
\end{equation}
Here we have $\nu_j\equiv N_j-N$ and $\kappa_j\equiv K_j-N_{j-1}$, with the restriction $\kappa_j-\nu_j>0$ for $j=1,\ldots,M$. The theta function in~\eqref{TruncUniDens} is included in the Meijer $G$-function, because the weight function~\eqref{wU} is strictly zero outside the unit disc. For example we have~\cite{Bateman:1953}:
\begin{equation}
\MeijerG{1}{0}{1}{1}{1}{0}{\abs z^2}=\Theta(1-\abs z^2)
\quad\text{and}\quad
\MeijerG{2}{0}{2}{2}{1,1}{0,0}{\abs z^2}=-\log\abs z^2\Theta(1-\abs z^2).
\end{equation}
Again the weight function can be written via an $M$-fold integral representation~\cite{ARRS,IK,ABKN},
\begin{multline}
w_{\nu,\mu}^{\beta=2}(z)=\pi^{1-M}
\prod_{m=1}^M\frac1{\Gamma(\kappa_m-\nu_m)}\,\int_D d^2z_m\abs{z_m}^{2\nu_m}(1-\abs{z_m}^2)^{\kappa_m-\nu_m-1}\\
\times\delta^2\left(z-z_M\cdots z_1\right),
\end{multline}
where the integration domain, $D$, is the unit disk. From this representation we find the squared norms
\begin{equation}
h_n^{\nu,\mu}=\pi\prod_{m=1}^M\frac{\Gamma(\nu_m+n+1)}{\Gamma(\kappa_m+n+1)}.
\label{hTruncComplex}
\end{equation}
The \emph{kernel} and the \emph{normalisation constant} are uniquely determined by~\eqref{KNdef} and~\eqref{ZNgeneral}, respectively.

In complete analogy to the Ginibre case we may go to the generalised eigenvalue problem~\eqref{charEq} and consider a \emph{mixed product of truncated and inverse truncated matrices}. We introduce $L$ additional independent unitary matrices $V_j$ of different sizes $T_j$ for $j=1,\ldots,L$, which are distributed uniformly with respect to Haar measure. Truncating each matrix $V_j$ to its upper-left sub-block $Y_j$ of size $N_{M+j-1}\times N_{M+j}$, the general eigenvalue problem gives rise to a jpdf with exactly the same structure as~\eqref{jpdf2}, except that the weight function~\eqref{wU} is replaced by~\cite{ARRS}
\begin{equation} 
w_{\nu,\mu,\kappa,\tau}^{\beta=2}(z)=\MeijerG{M}{L}{M+L}{M+L}{-N-\mu_1,\ldots,-N-\mu_L,\kappa_1,\ldots,\kappa_M}{\nu_1,\ldots,\nu_M,-N-\tau_1,\ldots,-N-\tau_L}{\abs z^2}.
\label{wUU}
\end{equation}
Here $\mu_j\equiv N_{M+j}-N$ and $\tau_j\equiv T_j-N_{M+j-1}$ satisfy the restriction $\tau_j-\mu_j>0$ for $j=1,\ldots,L$. Note that if $M=0$ then the weight function~\eqref{wUU} is strictly zero inside the unit disc. 

\funkybreak

\noindent
The result for $\beta=2$ can be extended to $\beta=4$ in a simple manner. The \emph{jpdf of complex eigenvalues} is of the same form as~\eqref{jpdf4} except that the weight function and the normalisation constant changes, see~\cite{IK}. The \emph{weight function} for the product of truncated unitary symplectic matrices can be expressed in terms of~\eqref{wU} as
\begin{equation}
w_{\nu,\kappa}^{\beta=4}(z)=w_{2\nu,2\kappa-1}^{\beta=2}(z).
\end{equation}
It follows from~\eqref{OP4} that
\begin{equation}
s_n=\pi\prod_{m=1}^M\frac{\Gamma(2\nu_m+n+1)}{\Gamma(2\kappa_m+n)}.
\end{equation}
This determines the skew-orthogonal polynomials~\eqref{skewOPgeneral} and therefore all correlation functions via~\eqref{Rkb=4}, with the kernel given by~\eqref{Knb=4}.

\subsubsection{Mixed products}

\noindent
Finally, we mention that no new techniques are required in order to study more complicated products constructed from different combinations of Ginibre, inverse Ginibre, truncated unitary as well as inverse truncated unitary matrices. Such products will have a jpdf with a structure similar to the examples above and the weight function can again be expressed as a Meijer $G$-function, albeit with more indices. In fact, products of Ginibre and truncated unitary matrices have previously been studied in~\cite{IK} for all $\beta$, although the most general results are restricted to $\beta=2,4$ due to the incompleteness of a generalised real Schur decomposition as discussed in section~\ref{EVginibre}. We stress again that all matrix ensembles described in this section are isotropic, which implies that the ordering of the matrices is irrelevant for all statistical properties of the eigenvalues, see~\cite{IK}.

% -----------------------------------------

\subsection{Singular Values}
\label{SVfinite}

\noindent
In this section, we seek the statistical properties of the singular values rather than of the complex eigenvalues for some random product matrix. Explicitly, we consider products of Ginibre, inverse Ginibre, and truncated unitary matrices. It turns out that all these ensembles are \emph{polynomial ensembles}~\cite{KS,Kuijlaars:2015} which are  special types of 
\emph{biorthogonal ensembles}~\cite{Muttalib:1995,Borodin}. For this reason we first recall a few general properties of polynomial ensembles. Let $x_n$ ($n=1,\ldots,N$) be a set of positive variables. We are interested in a \emph{generic jpdf} of the following form
\begin{equation}
\mathcal P_{\jpdf}^{\beta=2}(x_1,\ldots,x_{N})=
\frac1{\mathcal Z_{N}^{\beta=2}}\prod_{1\leq i<j\leq N}(x_j-x_i)\det_{1\leq \ell,k\leq N}\left[w^{\beta=2}_{k-1}(x_\ell)\right],
\label{jpdfSVgeneral}
\end{equation}
where $\{w^{\beta=2}_{k}(x)\}$ is a collection of weight functions on the positive half-line. The \emph{$k$-point correlation functions} are defined by~\cite{Mehta}, 
\begin{equation}
R_k^{\beta}(x_1,\ldots,x_k)\equiv \frac{N!}{(N-k)!}
\prod_{n=k+1}^N\int_0^\infty dx_n\, {\mathcal P}^{\beta}_{\jpdf}(x_1,\ldots,x_N).
\end{equation}
It follows from the biorthogonal structure of the jpdf~\eqref{jpdfSVgeneral} that the $k$-point correlation functions including the jpdf form a \emph{determinantal point process}
\begin{equation}
R_k^{\beta=2}(x_1,\ldots,x_k)
=\det_{1\leq i,j\leq k}\left[K_{N}^{\beta=2}(x_i,x_j)\right]\ ,
\label{RkSV}
\end{equation}
with \emph{kernel}
\begin{equation}
K_N^{\beta=2}(x,y)=\sum_{n=0}^{N-1}\frac{1}{h_n}p_n(x)\psi_n(y).
\label{kernelBi}
\end{equation}
The functions $p_n(x)$ and $\psi_n(x)$ must satisfy the \emph{biorthogonality relation}
\begin{equation}
\int_0^\infty dx\, p_k(x)\psi_\ell(x)=h_k\delta_{k,\ell}.
\label{OPbi}
\end{equation}
Due to the structure of the jpdf~\eqref{jpdfSVgeneral} each $p_n(x)$ is a monic polynomial.
%%%%%%%%%%%%linear span 
For the orthogonal functions  $\psi_n(x)$ we require that their linear span 
given by $\mbox{span}\{\psi_0(x),\ldots, \psi_{N-1}(x)\}$ agrees with that 
of the weight functions $w_n^{\beta=2}(x)$, $\mbox{span}\{w_0^{\beta=2}(x),\ldots,
w_{N-1}^{\beta=2}(x)\}$.
%, while each $\psi_n(x)$ belongs to the span of the weight functions %$w_k^{\beta=2}(x)$, with indices $k=0,\ldots,n$. We require that the linear %span of the weights is ``monic'' meaning that the coefficient of the weight %function with index $n$ is equal to unity, i.e. %$\psi_n(x)=w_n^{\beta=2}(x)+\cdots$. 
%
The biorthogonal functions as well as the squared norms $h_n$ are uniquely defined. 
Finally, the normalisation constant in the jpdf~\eqref{jpdfSVgeneral} can be written as the product following \cite{Mehta}
\begin{equation}
Z_N^{\beta=2}=N!\prod_{n=0}^{N-1}h_n.
\label{ZSVgeneral}
\end{equation}
Hence the normalisation is completely determined by the squared norms.

% -----------------------------------------

\subsubsection{Ginibre matrices}
\label{SVgauss}

\noindent
Now, we are ready to discuss \emph{products of rectangular Ginibre matrices}. We keep the notation from section~\ref{EVginibre} and consider a product matrix $\Pi_M$ given by~\eqref{prod} where each $X_j$ is an $(N+\nu_j)\times (N+\nu_{j-1})$ matrix distributed according to~\eqref{Ginibre}. A derivation of the jpdf for the singular values of the product matrix in the $\beta=2$ case was presented in~\cite{akw,aik} and explicitly uses the Itzykson--Zuber integration formula~\cite{IZ:1980}. The absence of similar integration formulae for $\beta=1$ and $\beta=4$ have, so far, restricted explicit calculations to $\beta=2$. For some asymptotic quantities this difficulty can be circumvented using supersymmetric techniques~\cite{Mario15}.

Let $x_j$ ($j=1,\ldots,N$) denote the squared singular values of the product matrix $\Pi_M$, then the \emph{jpdf for the singular values of the product of complex Ginibre matrices} is given by~\eqref{jpdfSVgeneral} with weight functions~\cite{akw,aik}
\begin{equation}
w^{\beta=2}_{\nu;k}(x)=\MeijerG{M}{0}{0}{M}{-}{\nu_1,\ldots,\nu_{M-1},\nu_M+k}{x}.
\label{wSVgauss}
\end{equation}
The biorthogonal functions corresponding to these weights were first obtained in~\cite{akw,aik}. It was shown that the \emph{biorthogonal functions} and the \emph{squared norms} are given by 
\begin{align}
p_n(x)&=\sum_{k=0}^n\frac{(-1)^{n+k}n!}{(n-k)!\,k!}\bigg[\prod_{m=1}^M\frac{\Gamma(\nu_m+n+1)}{\Gamma(\nu_m+k+1)}\bigg]x^k, \label{pGaussPoly}\\
\psi_n(x)&=\sum_{k=0}^n\frac{(-1)^{n+k}n!}{(n-k)!\,k!}\frac{\Gamma(\nu_M+n+1)}{\Gamma(\nu_M+k+1)} \nn\\
&\qquad\qquad\times\MeijerG{M}{0}{0}{M}{-}{\nu_1,\ldots,\nu_{M-1},\nu_M+k}{x}, \label{psiGaussPoly}
\end{align}
\begin{align}
h_n&=n!\prod_{m=1}^M\Gamma(\nu_m+n+1). \label{hGauss}
\end{align}
It is immediately seen from this representation that the biorthogonal functions are monic, but for further calculation it is often useful to rewrite them in terms of special functions. The polynomial $p_n(x)$ can be written either as a hypergeometric function or as a Meijer $G$-function, while $\psi_n(x)$ can be written as a single Meijer $G$-function. Explicitly, we have~\cite{akw,aik}
\begin{align}
p_n(x)&=(-1)^n\frac{h_n}{h_0}\hypergeometric{1}{M}{-n}{\nu_1+1,\ldots,\nu_M+1}{x} 
\nn \\
&=
-h_n\,\MeijerG{0}{1}{1}{M+1}{n+1}{-\nu_1,\ldots,-\nu_M,0}{x}, \label{pGaussG} \\
\psi_n(x)&=\MeijerG{M+1}{0}{1}{M+1}{-n}{0,\nu_1,\ldots,\nu_M}{x}. \label{psiGaussG}
\end{align}
This determines the normalisation via~\eqref{ZSVgeneral} and all correlations via~\eqref{RkSV} with kernel~\eqref{kernelBi}. Furthermore, the explicit formulation of the biorthogonal functions~\eqref{pGaussG} and~\eqref{psiGaussG} allows a double contour integral representation of the kernel~\cite{AZ},
\begin{align}
&K_N^{M,\nu}(x,y) \nn\\
&=\frac1{(2\pi i)^2}\int_{\mathcal C}du\oint_\Sigma dv \frac{x^uy^{-v-1}}{u-v} 
\frac{\Gamma(u-N+1)}{\Gamma(v-N+1)}\frac{\Gamma(u+1)}{\Gamma(v+1)}
\prod_{m=1}^M\frac{\Gamma(u+\nu_m+1)}{\Gamma(v+\nu_m+1)} \nn \\
&=\int_0^1du\, \MeijerG{0}{1}{1}{M+1}{N}{-\nu_1,\ldots,-\nu_M,0}{ux}
\MeijerG{M+1}{0}{1}{M+1}{-N}{0,\nu_1,\ldots,\nu_M}{uy}, \label{2intRep}
\end{align}
where $\mathcal C$ is a straight line from $-\frac12-i\infty$ to $-\frac12+i\infty$ while $\Sigma$ encloses $1,\ldots,N$ in the positive direction without any intersection with $\mathcal C$. This is seen by writing the biorthogonal functions~\eqref{pGaussG} and~\eqref{psiGaussG} as their integral representations, see~\eqref{Gdef}, and inserting them into the expression for kernel~\eqref{kernelBi}; the sum can be performed and the first line of~\eqref{2intRep} is obtained. In the second line we have displayed a second real integral representation from \cite{AZ}. It is already very reminiscent of
that large-$N$ asymptotic result as we will see in section~\ref{SVhard}.

It was pointed out in~\cite{AZ} that the polynomials~\eqref{pGaussPoly} satisfy a stricter condition than biorthogonality; they are multiple orthogonal polynomials of type II with respect to the weights $w_{\nu;0}^{\beta=2}(x),\ldots,w_{\nu;M-1}^{\beta=2}(x)$ (see~\cite{Ismail:2005,Kuijlaars:2010} for a discussion of multiple orthogonal polynomials). This means that
\begin{equation}
\int_0^\infty dz\, x^\ell p_n(x)w_{\nu;k}^{\beta=2}(x)=0\ ,
\label{multiple}
\end{equation}
for $k=0,\ldots,M-1$ and $\ell=0,\ldots,\big\lceil\tfrac{n-k}M\big\rceil-1$, where $\lceil x\rceil$ denote the ceiling function. This multiple orthogonality may be used to establish $M+2$ term recurrence relations for the biorthogonal functions:
\begin{align}
xp_n(x)&=p_{n+1}(x)+\sum_{m=0}^Ma_{m,n}p_{n-m}(x),&&
a_{m,n}=\int_0^\infty dx\, x\,p_n(x)\frac{\psi_{n-m}(x)}{h_{n-m}}, \nn \\
x\frac{\psi_n(x)}{h_n}&=\frac{\psi_{n-1}(x)}{h_{n-1}}+\sum_{m=0}^Mb_{m,n}\frac{\psi_{n+m}(x)}{h_{n+m}}, &&
b_{m,n}=\int_0^\infty dx\, x\,p_{n+m}(x)\frac{\psi_{n}(x)}{h_n}.
\label{recurrence}
\end{align}
These recurrence coefficients were explicitly calculated in~\cite{AZ} but will not be repeated here.

% -----------------------------------------

\subsubsection{Ginibre and inverse Ginibre matrices}
\label{SVinv}

\noindent
Like for the complex eigenvalues we will also consider a \emph{mixed product of Ginibre and inverse Ginibre matrices}; here we follow~\cite{PF14}. We seek the jpdf for the squared singular values of the matrix
\begin{equation}
Y_L^{-1}\cdots Y_1^{-1}X_M\cdots X_1,
\end{equation}
where $X_i$ and $Y_j$ are distributed according to the induced density~\eqref{induced} with indices $\nu_i$ and $\mu_j$, respectively. Recall that any rectangular structure of the product matrix can be incorporated by choosing $\nu_i$ ($i=1,\ldots,M$) and  $\mu_j$ ($j=1,\ldots,L$) to be positive integers. 
The \emph{jpdf for the mixed product of Ginibre and inverse Ginibre matrices} is given by~\eqref{jpdfSVgeneral} with \emph{weight function}~\cite{PF14}
\begin{equation}
w^{\beta=2}_{\nu,\mu;k}(x)=\MeijerG{M}{L}{L}{M}{-N-\mu_1,\ldots,-N-\mu_L}{\nu_1,\ldots,\nu_{M-1},\nu_M+k}{x}.
\label{wSVinv}
\end{equation}
This choice of weight functions requires that $M\geq 1$ but allows $L=0$. However, there exists an alternative choice which requires that $L\geq 1$ and allows $M=0$, leading to the same correlation functions.

Similar to the pure Ginibre case, the biorthogonal functions may be expressed neatly in terms of special functions~\cite{PF14},
\begin{align}
p_n(x)&=(-1)^n\frac{h_n}{h_0} \hypergeometric{L+1}{M}{-n,1-N-\mu_1,\ldots,1-N-\mu_L}{\nu_1+1,\ldots,\nu_M+1}{(-1)^Lx} \nn \\
&=-h_n\MeijerG{0}{1}{L+1}{M+1}{n+1,N+\mu_1,\ldots,N+\mu_L}{-\nu_1,\ldots,-\nu_M,0}{x}, \label{pInv} \\
\psi_n(x)&=\MeijerG{M+1}{L}{L+1}{M+1}{-N-\mu_1,\ldots,-N-\mu_L,-n}{0,\nu_1,\ldots,\nu_M}{x}, \label{psiInv} \\
h_n&=n!\prod_{m=1}^M\Gamma(n+\nu_m+1)\prod_{\ell=1}^L\Gamma(N+\mu_\ell-n) \label{hInv}.
\end{align}
If $L=0$ then the biorthogonal functions reduce to the Ginibre case and they satisfy the $M+2$ term recurrence relations~\eqref{recurrence}. However, if $L\geq 1$ then the multiple orthogonality relations~\eqref{multiple} are no longer valid and no recurrence relations are known to date.

As in the Ginibre case, the normalisation constant and the $k$-point correlations are determined by the biorthogonal functions~\eqref{pInv} and~\eqref{psiInv} together with the squared norms~\eqref{hInv} via~\eqref{ZSVgeneral} and~\eqref{RkSV}, respectively. Also a double contour integral representation is possible,
\begin{multline}
K_N^{M,L,\nu,\mu}(x,y)=\frac1{(2\pi i)^2}\int_{-\frac12-i\infty}^{-\frac12+i\infty}\!du\oint_\Sigma dv \frac{x^uy^{-v-1}}{u-v}
\frac{\Gamma(u-N+1)}{\Gamma(v-N+1)}\frac{\Gamma(u+1)}{\Gamma(v+1)} \\
\times\prod_{m=1}^M\frac{\Gamma(u+\nu_m+1)}{\Gamma(v+\nu_m+1)}\prod_{\ell=1}^L\frac{\Gamma(N+\mu_m-u)}{\Gamma(N+\mu_m-v)}.
\end{multline}
The contour $\Sigma$ is defined as for products of Ginibre matrices and a similar single real integral representation as in~\eqref{2intRep} exists.

\subsubsection{Truncated unitary matrices}
\label{SVtrunc}

\noindent
We now turn to \emph{products of truncated unitary matrices}. The derivation of the jpdf for such products requires an extension of the Itzykson--Zuber integral which was very recently proved in~\cite{KKS}. Moreover, the authors of~\cite{KKS} showed that the \emph{jpdf for the product of $M$ truncated unitary matrices} was given by~\eqref{jpdfSVgeneral} with \emph{weight function}
\begin{equation}
w^{\beta=2}_{\nu,\kappa;k}(x)=\MeijerG{M}{0}{M}{M}{\kappa_1,\ldots,\kappa_{M-1},\kappa_M-N+k+1}{\nu_1,\ldots,\nu_{M-1},\nu_M+k}{x}.
\label{wSVtrunc}
\end{equation}
Here we use the same notation as in section~\ref{EVtrunc}. The biorthogonal functions are given by
\begin{align}
p_n(x)
&=(-1)^n\frac{h_n}{h_0} \hypergeometric{M+1}{M}{-n,\kappa_1+1,\ldots,\kappa_M+1}{\nu_1+1,\ldots,\nu_M+1}{x}  \nn \\
&=h_n\,\MeijerG{0}{M+1}{M+1}{M+1}{-\kappa_1,\ldots,-\kappa_M,n+1}{0,-\nu_1,\ldots,-\nu_M}{x}, \label{pTrunc} \\
\psi_n(x)&=\MeijerG{M+1}{0}{M+1}{M+1}{-n,\kappa_1,\ldots,\kappa_M}{\nu_1,\ldots,\nu_M,0}{x}, \label{psitrunc} \\
h_n&=n!\prod_{m=1}^M\frac{\Gamma(\nu_m+n+1)}{\Gamma(\kappa_m+n+1)} \label{hTrunc},
\end{align}
which determines the normalisation~\eqref{ZSVgeneral} and the correlations~\eqref{RkSV}. Similar to the two previous examples, the kernel can be written as a double contour integral,
\begin{multline}
K_N^{M,\nu,\kappa}(x,y)=\frac1{(2\pi i)^2}\int_{-\frac12-i\infty}^{-\frac12+i\infty}du\oint_\Sigma dv \frac{x^uy^{-v-1}}{u-v}
\frac{\Gamma(u-N+1)}{\Gamma(v-N+1)}\frac{\Gamma(u+1)}{\Gamma(v+1)} \\
\times\prod_{m=1}^M\frac{\Gamma(u+\nu_m+1)}{\Gamma(v+\nu_m+1)}\frac{\Gamma(v+\kappa_m+1)}{\Gamma(u+\kappa_m+1)},
\end{multline}
where the contour $\Sigma$ is chosen such that it encloses $1,\dots,N$ without intersecting the other contour. For the real integral representation corresponding to~\eqref{2intRep} we refer to \cite{KKS}.

\subsubsection{Mixed products}
\label{SVmix}

\noindent
As a final remark, we mention that more complicated products constructed from Ginibre, inverse Ginibre, truncated unitary and inverse truncated unitary matrices can be studied without introducing new techniques. Such a product of Ginibre matrices times a single truncated unitary matrix has previously been studied in~\cite{KS}. When mixed products are considered it is important to note that all matrix ensembles described in this section are isotropic, which implies that the ordering of the matrices is irrelevant for the statistical properties of the singular values, see~\cite{IK}.

\section{Local universality for large matrix dimensions}
\label{largeN}

\noindent
In this section we will discuss the universal limits for products of a finite number $M$ of matrices as the matrix dimension $N$ tends to infinity. Global spectra for both the complex eigenvalues and the singular values of product matrices have been studied intensively in the literature; in particular using techniques from free probability, see~\cite{GT,GKT:2014,AGT:2010,BBCC:2011,BJW,Burda,NS:2006,OS,ORSV,Karol}. Many of these results predate the exactly solvable cases described in section~\ref{finiteNM}. However, almost nothing was known about the local universality for products of random matrices until recently. One of the main benefits of the matrix products described in section~\ref{finiteNM} is that their exact solvability allows a direct study of both global and local universality. Here we will restrict our attention to results about local universality. In the known cases, it turns out that local correlations in the bulk and at the soft edges correspond to classical universality results from random 
matrix 
theory; that is Ginibre-type correlations for the complex eigenvalues and correlations given in terms of the sine and Airy kernels for the singular values. At the origin and at the hard edge new universality classes arise. In order to make this review both short and concise, our main focus will be on these new universality classes. We stress that this choice does not mean that the results for the bulk and the soft edges are less interesting, neither that these results are easily obtained.

\subsection{Complex eigenvalues}
\label{EVlocal}

\subsubsection{Origin}

\noindent
We consider the local correlations at the origin for a product of complex ($\beta=2$) Ginibre matrices first obtained for square matrices in~\cite{ab}. The correlations are given by~\eqref{Rkb=2} with weight~\eqref{wGdef} and kernel~\eqref{Kgauss2}. The local scale at the origin is obtained by keeping the number of matrices $M$ as well as the parameters $\nu_m>-1$ ($m=1,\ldots,M$) fixed, while taking the matrix dimension $N$ to infinity without further rescaling. This is a trivial task since the weight function~\eqref{wGdef} is independent of $N$, while the kernel~\eqref{Kgauss2} simply becomes an infinite sum which can be written either as a hypergeometric function or a Meijer $G$-function~\cite{ab},
\begin{align}
K_\origin^{M,\nu}(u,v)&\equiv\lim_{N\to\infty}K_{N,\nu}^{\beta=2}(u,v)
=\frac1\pi \sum_{n=0}^\infty \frac{(uv^*)^n}{\prod_{m=1}^M\Gamma(\nu_m+n+1)} \nn\\
&=\frac{1}{\pi}\frac{\hypergeometric{1}{M}{1}{\nu_1+1,\ldots,\nu_M+1}{uv^*}}{\prod_{m=1}^M\Gamma(\nu_m+1)} \nn \\
&=\frac1\pi\,\MeijerG{1}{1}{1}{M+1}{0}{0,-\nu_1,\ldots,-\nu_M}{-uv^*}.
\label{Korigin}
\end{align}
The \emph{universal $k$-point correlation functions at the origin} therefore read
\begin{equation}
\rho^{M,\nu}_{\origin;k}(z_1,\ldots,z_k)=\prod_{\ell=1}^k w^{M}_{\nu}(z_\ell)\det_{1\leq i,j\leq k}\left[K_\origin^{M,\nu}(z_i,z_j)\right],
\label{Rorigin}
\end{equation}
with kernel~\eqref{Korigin} and weight
\begin{equation}
w^{M}_{\nu}(z)=\MeijerG{M}{0}{0}{M}{-}{\nu_1,\ldots,\nu_M}{\abs z^2}.
\label{wOrigin}
\end{equation}
For $M=1$ and $\nu=0$, the product ensemble reduces to the classical Ginibre ensemble. Evaluating the special functions~\eqref{Korigin} and~\eqref{wOrigin} for $M=1$ and inserting this into~\eqref{Rorigin}, we explicitly see that
\begin{align}
\label{ob}
\rho^{M=1,\nu=0}_{\origin;k}(z_1,\ldots,z_k)&=\det_{1\leq i,j\leq k}\left[\frac1\pi \exp\left(-\tfrac12\abs{z_i}^2-\tfrac12\abs{z_j}^2+z_iz_j^*\right)\right],\\
&= \rho^{\beta=2}_{\bulk;k}(z_1,\ldots,z_k). \nn
\end{align}
These are the known correlation functions for a single complex Ginibre matrix at the origin. Because in the Ginibre ensemble the density is flat the origin is not special and eq.\eqref{ob} agrees with the bulk scaling limit, see the next subsection \ref{bse}.

For $M\geq2$, the correlations~\eqref{Rorigin} satisfy a reduction relation
\begin{equation}
\lim_{\nu_M\to\infty}{(\nu_M)^k}\rho^{M,\nu}_{\origin;k}(\sqrt{\nu_m} z_1,\ldots,\sqrt{\nu_M}z_k)=\rho^{M-1,\nu}_{\origin;k}(z_1,\ldots,z_k).
\end{equation}
We recall that the parameters $\nu_m$ ($m=1,\ldots,M$) incorporate the rectangular structure of the matrices.

Let us now turn to a mixed product of Ginibre and inverse Ginibre matrices as discussed in section~\ref{EVinv}. Here, the $k$-point correlation functions are given by~\eqref{Rkb=2} with weight~\eqref{wGG} and kernel~\eqref{KNdef}, where the squared norms are given by~\eqref{moments2}. It can be seen that with $\nu_m$ and $\mu_m$ fixed for $m=1,\ldots,M$, we have the following scaling limit for the $k$-point correlation function:
\begin{equation}
\lim_{N\to\infty}\frac1{N^{kL}}R^{M,L,\nu,\mu}_{\origin;k}\bigg(\frac{z_1}{N^{L/2}},\ldots,\frac{z_k}{N^{L/2}}\bigg)
=\rho^{M,\nu}_{\origin;k}(z_1,\ldots,z_k).
\end{equation}
Here the right hand side is given by~\eqref{Rorigin}, i.e. it is the same as for the product without any inverse matrices $L=0$.

Finally, we will look at products of truncated unitary matrices. We consider the weight function~\eqref{wU} and the kernel~\eqref{KNdef} with the squared norms given by~\eqref{hTruncComplex}. Let $J$ and $L$ be integers such that $J+L=M$. We take $M$ and $J$ as well as $\nu_m$ ($m=1,\ldots,M$) and $\kappa_j$ ($j=1,\ldots,J$) to be fixed, while $\kappa_\ell=N+O(1)$ as $N$ tends to infinity for $\ell=1,\ldots,L$. The $k$-point correlation functions~\eqref{Rkb=2} have a hard edge scaling limit given by
\begin{align}
\rho^{M,J,\nu,\kappa}_{\origin;k}(z_1,\ldots,z_k)
&\equiv\lim_{N\to\infty}\frac1{N^{kL}}R^{M,J,\nu,\kappa}_{\origin;k}\bigg(\frac{z_1}{N^{L/2}},\ldots,\frac{z_k}{N^{L/2}}\bigg) \nn \\
&=\prod_{\ell=1}^k w^{M,J}_{\nu,\kappa}(z_\ell)\det_{1\leq i,j\leq k}\left[K_\origin^{M,J,\nu,\kappa}(z_i,z_j)\right],
\label{MeijerComplex}
\end{align}
where the weight and kernel are given by
\begin{align}
w^{M,J}_{\nu,\kappa}(z)&=\MeijerG{M}{0}{J}{M}{\kappa_1,\ldots,\kappa_J}{\nu_1,\ldots,\nu_M}{\abs z^2}, \\
K_\origin^{M,J,\nu,\kappa}(u,v)
&=\frac{1}{\pi}\frac{\hypergeometric{J+1}{M}{1,\kappa_1+1,\kappa_J+1}{\nu_1+1,\ldots,\nu_M+1}{uv^*}}{\prod_{m=1}^M\Gamma(\nu_m+1)/\prod_{j=1}^J\Gamma(\kappa_j+1)} \nn \\
&=\frac1\pi\,\MeijerG{1}{J+1}{J+1}{M+1}{0,-\kappa_1,\ldots,-\kappa_J}{0,-\nu_1,\ldots,-\nu_M}{-uv^*}.
\end{align}
This was first shown in~\cite{ABKN} for a product with $J=0$; in this case the correlation functions~\eqref{MeijerComplex} reduce to the case with $M$ Ginibre matrices~\eqref{Rorigin}. It remains to find the local correlations  for more general products. It would be natural to look at mixed products constructed from both Ginibre and truncated unitary matrices as well as their inverses.

\funkybreak

\noindent
A more challenging task is to find the local correlations at the origin for products of real ($\beta=1$) and quaternionic ($\beta=4$) matrices.
For $\beta=4$, the correlations at the origin were obtained for a single matrix as well as a product of two matrices in~\cite{Kanzieper:2002} and~\cite{a05}, respectively. The main idea presented in~\cite{Kanzieper:2002} and reused in~\cite{a05} was to write down a set of coupled differential equations for the kernel at the origin. This technique can be extended to products of an arbitrary number of matrices. However, the complicated structure the differential equations for arbitrary $M$ has prevented an explicit evaluation for $M\geq3$, so far. The calculations simplify considerably if we integrate out the angular dependence of the eigenvalues; this idea was explicitly used in~\cite{Jesper,IK,ais}. The local radial density at the origin for products of Ginibre matrices was explicitly calculated in~\cite{Jesper} and a structure closely related that of complex matrices was found. For $\beta=1$ the correlations for $M=2$ were found in~\cite{aps}, but almost nothing is known for $M\geq3$.

\subsubsection{Bulk and soft edge}\label{bse}

\noindent
It was shown in~\cite{ab,ABKN,LW} that, under proper rescaling, the local \emph{$k$-point correlation functions of the eigenvalues in the bulk} for a product of complex Ginibre or truncated unitary matrices are given by
\begin{equation}
\rho^{\beta=2}_{\bulk;k}(z_1,\ldots,z_k)=\det_{1\leq i,j\leq k}\Big[\frac1\pi \exp\big(-\tfrac12\abs{z_i}^2-\tfrac12\abs{z_j}^2+z_iz_j^*\big)\Big],
\end{equation}
which is the same as for the standard complex Ginibre ensemble, see e.g.~\cite{Mehta,Forrester:2010}. In order to study the local correlations in the vicinity of a (soft) edge, we need to choose a point located on the edge. Given such a point, $z_0$, it was shown~\cite{ab,ABKN,LW} that proper rescaling leads to
\begin{multline}
\rho^{\beta=2}_{\soft;k}(z_1,\ldots,z_k)=\\
\det_{1\leq i,j\leq k}\bigg[\frac1{2\pi} \exp\big(-\tfrac12\abs{z_i}^2-\tfrac12\abs{z_j}^2+z_iz_j^*\big)\erfc\bigg(\frac{z_0^*z_i+z_j^*z_0}{\sqrt 2}\bigg)\bigg],
\end{multline}
where $\erfc(x)$ is the complementary error function. Note that this supplements the universality results at strong non-Hermiticity known for the complex eigenvalues of non-Hermitian matrices~\cite{AHM:2011,Berman:2008,TV:2012}, see also~\cite{PP:2014} for a recent heuristic approach.

%%%%%%%%%%%%%%%%%%%%%%%%%%%%%%%%

\subsubsection{Weak non-unitarity limit}

\noindent
In the case of truncated unitary (or orthogonal) matrices one may consider the particular limit, in which the number of truncations remains finite while the matrix size(s) go to infinity. Consequently the resulting truncated matrices become almost unitary, with the macroscopic density of complex eigenvalues condensing on the unit circle. However, locally the complex eigenvalues may still extend inside the unit disc. For $M=1$ this limit was studied first in \cite{KS} and named weakly non-unitary. For $M>1$ and a particular choice of parameters this has be generalised in \cite{ABKN}. Namely if in subsection 
\ref{EVtrunc} we truncate all matrices in the product starting from the same size, $K_j=K$, down to square matrices of size $N_j=N$ ($\nu_j=0$) for all $j$, 
we take the following large-$N$ limit with $K-N=\kappa$ fixed.
In order obtain non-trivial local correlations inside the unit disc we take $k$ points $z_l$ in the vicinity of a fixed point $z_0$ on the unit circle:
\begin{equation}
z_j=1-\frac1N (x_j+iy_j)\ ,\ x_j>0 \ \ \mbox{for} \ \ j=1,2,\ldots,k.
\end{equation}
Here we have chosen $z_0=1$ without loss of generality, due to the rotation invariance. In contrast to the previous sections here the weight \eqref{wU} and the kernel \eqref{KNdef} with norms \eqref{hTruncComplex} do not converge to a limit individually. Only their combination as it appears in the correlation functions \eqref{Rkb=2} has a limit which we give here straight away:
\begin{multline}
\lim_{N\to\infty}\frac{1}{N^{2k}}R_k^{\beta=2}\left(1-\frac1N (x_1+iy_1),\ldots,
1-\frac1N (x_k+iy_k)\right)\\
=\det_{1\leq j,l\leq k}\left[
K_{\rm weak}^{M,\kappa}(x_j+iy_j,x_l+iy_l)\right],
\end{multline}
where
the \emph{weak kernel} is given by~\cite{ABKN}
\begin{align}
K_{\rm weak}^{M,\kappa}(x_j+iy_j,x_l+iy_l)&=\frac{\Theta(x_j)\Theta(x_l)
(4x_jx_l)^{(M\kappa-1)/2}}{\pi(M\kappa-1)!}\\
&\times\left(-\frac{\partial}{\partial t}\right)^{M\kappa}
\left.\left(\frac{1-e^{-t}}{t}\right)\right|_{t=(x_j+x_l+i(y_j-y_l))/M}.\nn
\end{align}
This expression generalises the kernel at weak non-unitarity given in \cite{KS} for $M=1$.
In order to extend this result to a more general parameter setting with the $\kappa_j$ being different, the difficulty is to obtain the limiting kernel (and not the weight).

\subsubsection{Further limits}

\noindent
In addition to the various limits in different parts of the spectrum we discussed so far further asymptotic limits can be considered. For example the large radius limit $r\to\infty$ of the hole probabilities eqs. \eqref{holeb2} and \eqref{holeb4} at finite and infinite $N$ 
have been derived in \cite{as} and \cite{ais} for $\beta=2$ and 4, respectively. 
Furthermore the infinite-$N$ process can be considered directly and for example overcrowding estimates can be made  \cite{as,ais}.
Similar questions arise in the analysis of zeros of Gaussian analytic functions, cf. \cite{Hough}.

A related question was considered in \cite{JQ}, where the distribution of the largest radius was studied in a double scaling limit with both $N$ and $M$ going to infinity. In the case of products of square Ginibre matrices with $\beta=2$ a transition between the known Gumbel distribution for $M=1$ \cite{Rider03} and a lognormal distribution was found. The latter relates to the study of the largest stability exponent which will be introduced in section \ref{largeM}.

% ------------------------------------------------------

\subsection{Singular values}
\label{SVasymp}

\subsubsection{Hard edge}
\label{SVhard}

\noindent
In this section we return to statistical properties of the squared singular values for products of random matrices. The products described in section~\ref{SVfinite} were all determinantal point processes~\eqref{RkSV} with a kernel~\eqref{kernelBi} constructed from a set of biorthognal functions. A study of the local correlations for such products at the hard edge for large matrix dimensions was first undertaken in~\cite{AZ}, where it was found that a new family of universal correlation kernels arises. 
We saw in section~\ref{SVgauss} that the kernel for the product of Ginibre matrices can be written as a double contour integral~\eqref{2intRep}. Using that
\begin{equation}
\frac{\Gamma(u-N+1)}{\Gamma(v-N+1)}=\frac{\sin\pi u}{\sin\pi v}N^{u-v}(1+O(N^{-1})),
\end{equation}
we find the hard edge limit
\begin{align}
K_\meijer^{M,\nu}(x,y)&\equiv\lim_{N\to\infty}\frac{1}{N}K_N^{M,\nu}\Big(\frac xN,\frac yN\Big) \nn \\
&=\frac1{(2\pi i)^2}\int_{-\frac12-i\infty}^{-\frac12+i\infty}du\oint_\Sigma dv \frac{x^uy^{-v-1}}{u-v}
\frac{\sin\pi u}{\sin\pi v}\frac{\Gamma(u+1)}{\Gamma(v+1)} \nn \\
&\qquad\qquad\qquad\qquad\qquad\qquad\times\prod_{m=1}^M\frac{\Gamma(u+\nu_m+1)}{\Gamma(v+\nu_m+1)}.
\end{align}
Evaluating the integrals allows a representation in terms of special functions
\begin{align}
&K_\meijer^{M,\nu}(x,y) \nn\\
&\equiv\int_0^1ds\,\frac{\hypergeometric{0}{M}{-}{\nu_1+1,\ldots,\nu_M+1}{-sx}}{\prod_{m=1}^M\Gamma(\nu_m+1)}\MeijerG{M}{0}{0}{M+1}{-}{\nu_1,\ldots,\nu_M,0}{sy} \nn \\
&=\int_0^1ds\,\MeijerG{1}{0}{0}{M+1}{-}{0,-\nu_1,\ldots,-\nu_M}{sx}\MeijerG{M}{0}{0}{M+1}{-}{\nu_1,\ldots,\nu_M,0}{sy}.
\label{MeijerKernel}
\end{align}
We will refer to~\eqref{MeijerKernel} as the \emph{Meijer $G$-kernel}~\cite{KKS}. The Hamilton equations for the gab probability in this limit were studied in~\cite{ES14}.

For $M=1$ the product consists of a single matrix, hence it reduces to the Wishart--Laguerre ensemble. It is well-known that the hard edge behaviour for the Wishart--Laguerre ensemble is described by the Bessel kernel, see e.g.~\cite{Forrester:2010}. This is explicitly incorporated in the Meijer $G$-kernel~\eqref{MeijerKernel} by
\begin{equation}
K_\meijer^{M=1,\nu}(x,y)=\Big(\frac yx\Big)^{\nu/2}\!\!\int_0^1\!\!ds J_\nu(2\sqrt{ux})J_\nu(2\sqrt{uy})
=4\Big(\frac yx\Big)^{\nu/2}K_\bessel^{\nu}(4x,4y), \label{MeijerBessel}
\end{equation}
where the \emph{Bessel kernel} is defined as
\begin{equation}
K_\bessel^{\nu}(x,y)\equiv\frac{\sqrt y J_\nu(\sqrt x)J_\nu'(\sqrt y)-\sqrt x J_\nu'(\sqrt x)J_\nu(\sqrt y)}{2(x-y)}.
\end{equation}
Note that the $x$ and $y$ dependent prefactor to the Bessel kernel in~\eqref{MeijerBessel} cancels out when calculating the correlation functions due to the determinantal structure.
For $M\geq 2$, the Meijer $G$-kernel~\eqref{MeijerKernel} satisfies a reduction relation,
\begin{equation}
\lim_{\nu_M\to\infty}{\nu_M}K_N^{M,\nu}({\nu_M}x,{\nu_M}y)=K_\meijer^{M-1,\nu}(x,y),
\end{equation}
where the kernel on the right hand side depends on the remaining parameters $\nu_m$ ($m=1,\ldots,M-1$). Recall that for the product of rectangular Ginibre matrices, the parameters $\nu_m$ incorporate the rectangular structure.

Similar results were obtained for mixed products of Ginibre and inverse Ginibre matrices in~\cite{PF14} and for products of truncated unitary matrices in~\cite{KKS}; the product of Ginibre matrices together with a single truncated unitary matrix was studied in~\cite{KS}. The biorthognal functions for mixed products of Ginibre and inverse Ginibre matrices are given in section~\ref{SVgauss}. For $M$ and $L$ as well as $\nu_m$ ($m=1,\ldots,M$) and $\mu_m$ ($\ell=1,\ldots,L$) fixed, we have~\cite{PF14}
\begin{equation}
\lim_{N\to\infty}\frac{1}{N^{L+1}}K_N^{M,L,\nu,\mu}\Big(\frac x{N^{L+1}},\frac y{N^{L+1}}\Big)=K_\meijer^{M,\nu}(x,y),
\end{equation}
where the kernel on the left hand side is given by~\eqref{kernelBi}.

For products of truncated unitary matrices the biorthogonal functions as well as a double contour integral representation of the kernel can be found in section~\ref{SVtrunc}.  Let $J$ and $L$ be integers such that $J+L=M$. We take $M$ and $J$ as well as $\nu_m$ ($m=1,\ldots,M$) and $\kappa_j$ ($j=1,\ldots,J$) to be fixed, while $\kappa_\ell=N+O(1)$ as $N$ tends to infinity for $\ell=1,\ldots,L$. In this case, we find
\begin{equation}
\lim_{N\to\infty}\frac{1}{N^{L+1}}K_N^{M,J,\nu,\kappa}\Big(\frac x{N^{L+1}},\frac y{N^{L+1}}\Big)=K_\meijer^{M,J,\nu,\kappa}(x,y),
\end{equation}
where the kernel on the right hand side is a generalised version of the kernel~\eqref{MeijerKernel} given by~\cite{KKS}
\begin{align}
&K_\meijer^{M,J,\nu,\kappa}(x,y) \nn \\
&=\int_0^1ds\,\frac{\hypergeometric{J}{M}{\kappa_1+1,\ldots,\kappa_J+1}{\nu_1+1,\ldots,\nu_M+1}{-sx}}%
   {\prod_{m=1}^M\Gamma(\nu_m+1)/\prod_{j=1}^J\Gamma(\kappa_j+1)}%
   \MeijerG{M}{0}{J}{M+1}{\kappa_1,\ldots,\kappa_J}{\nu_1,\ldots,\nu_M,0}{sy} \nn \\
&=\int_0^1ds\,\MeijerG{1}{J}{J}{M+1}{-\kappa_1,\ldots,-\kappa_J}{0,-\nu_1,\ldots,-\nu_M}{sx}%
   \MeijerG{M}{0}{J}{M+1}{\kappa_1,\ldots,\kappa_J}{\nu_1,\ldots,\nu_M,0}{sy}.
\end{align}
Note that this Meijer $G$-kernel reduces to~\eqref{MeijerKernel} for $J=0$, exactly like the $k$-point correlation functions~\eqref{MeijerComplex} reduce to~\eqref{Rorigin} for $J=0$.

In addition to the product ensembles described above, the Meijer $G$-kernel~\eqref{MeijerKernel} has appeared in the context of Cauchy multi-matrix models~\cite{MB2,BGS:2014,MB1,FK:2014} and Muttalib--Borodin ensembles~\cite{Borodin,Muttalib:1995}. The reappearance the of the Meijer $G$-kernel~\eqref{MeijerKernel} in all these models suggests a much stronger underlying universality principle similar to known results for the Bessel kernel~\cite{ADMN:1997,KF:1998,KV:2002,KV:2003,Lubinsky:2008}.
In the global regime such a link is provided by the equilibrium measure~\cite{FL}.

\subsubsection{Bulk and soft edge}

\noindent
The evaluation of the local statistics in the bulk and at the soft edge turns out to be much more technically demanding than the evaluation at the hard edge, and has only very recently been obtained for the product of rectangular Ginibre matrices~\cite{lundang}. There it was shown that classical random statistics is obtained under proper rescaling, i.e. the \emph{sine kernel} in the bulk and the \emph{Airy kernel} at the soft edge. Recall that the sine and Airy kernel is defined by (see e.g.~\cite{Mehta})
\begin{equation}
K_\sine(x,y)\equiv \frac{\sin\pi(x-y)}{x-y}
\ \ \text{and}\ \ 
K_\airy(x,y)\equiv \frac{\Ai(x)\Ai'(y)-\Ai'(x)\Ai(y)}{x-y},
\end{equation}
respectively. We emphasise that the derivation of the sine and Airy kernel presented in~\cite{lundang} crucially depends on known results for the global density of the squared singular values~\cite{BBCC:2011,LSW:2011,Karol}, and in particular an asymptotic expression in terms of elementary functions obtained in~\cite{Biane:1998,HM:2013,TN}. However, the strategy for obtaining the local statistics presented in~\cite{lundang} is by no means restricted to the Ginibre case; in fact, Ref. \cite{lundang} provides similar results for mixed products of Ginibre and inverse Ginibre matrices (see section~\ref{SVinv}) and for Ginibre matrices mixed with a single truncated unitary matrix (see~\cite{KS}) although the proofs are only sketched.

\section{Lyapunov and stability exponents for large products}
\label{largeM}

\noindent
Until now we have considered the asymptotic behaviour of matrix products with a finite number of factors $M$ as the matrix dimensions tend to infinity. In this section we will consider the opposite situation, where the matrix dimensions are kept fixed as the number of factors $M$ tends to infinity. In general, we are interested in the spectral properties of a product matrix
\begin{equation}
\Pi_M\equiv X_M\cdots X_1,
\label{prod2}
\end{equation}
where each $X_m$ ($m=1,\ldots,M$) is an $N\times N$ random matrix, independently chosen from some ensemble. The multiplicative ergodic theorem of Oseledec~\cite{Oseledec:1968,Raghunathan:1979} states that if the second moments of the diagonal entries of $X_m^\dagger X_m$ are finite, then there is a well-defined limiting matrix
\begin{equation}
V\equiv \lim_{M\to\infty} (\Pi_M^\dagger \Pi_M)^{1/2M}
\end{equation}
with real eigenvalues $e^{\lambda_n}$ ($n=1,\ldots,N$). Here, the $\lambda_n$ are known as Lyapunov exponents. Let $x_n$ ($n=1,\ldots,N$) denote the squared singular values of the product matrix~\eqref{prod2}; Oseledec's theorem tells us that $x_n\sim e^{2M\lambda_n}$ for large $M$, hence negative (positive) Lyapunov exponents represent exponential decay (growth) and therefore stability (instability). Moreover, we see that all squared singular values diverge exponentially whenever the Lyapunov spectrum is non-degenerate. In such cases, it is expected that the Lyapunov exponents become independent Gaussian random variables with fluctuations of order $M^{-1/2}$ for sufficiently large $M$, see~\cite{CKN:1986,CPV,Virster:1970}. 
Together with the symmetry of the Lyapunov exponents under permutations this statement is equivalent to saying that
the jpdf for the squared singular values is given by a permanental point process,
\begin{multline}
\prod_{n=1}^N 2Me^{2M\xi_n}\,\mathcal P_\jpdf(e^{2M\xi_1},\ldots,e^{2M\xi_N})\\
\sim N!\per_{1\leq i,j\leq N}\left[\sqrt{\frac{M}{2\pi\sigma_i^2}}\exp\bigg(-\frac{M(\xi_j-\lambda_i)^2}{2\sigma_i^2}\bigg)\right],
\label{LyapunovPer}
\end{multline}
with means $\lambda_i$ and variances $\sigma_i$ to be determined.
The Gaussian approximation is expected to be valid when the distance between Lyapunov exponents is large compared to the size of their fluctuations, i.e. $\abs{\lambda_i-\lambda_j}\gg M^{-1/2}$ for all $i\neq j$.

How much of this expectation has been realised? In~\cite{CN:1984}, the mean and variance of the largest Lyapunov exponent has been computed for products of real Ginibre matrices, see~\eqref{Lgauss} below. The same ideas were used to calculate all the Lyapunov exponents in~\cite{Newman:1986}. In these papers probabilistic methods were applied, which we will briefly recall below.
Further explicit results for the Lyapunov exponents (again based on probability theory) were derived much later for factors from correlated Gaussian ensembles in~\cite{PF13} for $\beta=2$ and~\cite{Kargin} for $\beta=1,4$, see $\lambda_i$ in~\eqref{Lgauss} for the uncorrelated case.

It is clear from the discussion in section~\ref{SVfinite} that the explicit results for finite-$N$ and -$M$ should give access to the permanental structure.
Indeed in~\cite{ABK} such a derivation was recently performed for the product of Ginibre matrices with $\beta=2$. There the permanental point process~\eqref{LyapunovPer} was derived starting from \eqref{jpdfSVgeneral}, defining $x_j=e^{2M\xi_j}$ and taking $M$ to be large.
The following values for the variances (and means) were obtained for $\beta=2$:
\begin{equation}
\lambda_n^\beta=\frac12\log\frac2\beta+\frac12\psi\bigg(\frac{\beta n}2\bigg)
\quad\text{and}\quad
(\sigma_n^\beta)^2=\frac14\psi'\bigg(\frac{\beta n}2\bigg),
\label{Lgauss}
\end{equation}
where $\psi(x)$ denotes the digamma function. For further details including the subleading higher order cumulants for $\beta=2$ we refer to~\cite{ABK}.

Previously a direct access to the jpdf at finite $M$ and $N$ was unknown, and alternative methods for the evaluation of the Lyapunov exponents were developed much earlier~\cite{CN:1984}. These are 
valid for the more general class of isotropic matrices (including the Ginibre case)  which is why we recall them here.
The partial sum of Lyapunov exponents can be written as
\begin{equation}
\Lambda_k^\beta\equiv\sum_{n=1}^k\lambda_{N-n+1}^\beta\stackrel{d}{=}\frac1{2M}\sum_{m=1}^M\sup_{A\in \mathbb F^{N\times k}}\log\frac{\det A^\dagger X_m^\dagger X_m A}{\det A^\dagger A},
\label{newman}
\end{equation}
where $\lambda^\beta_{N-n+1}$ denotes the $n$-th largest Lyapunov exponent and the supremum is over all $N\times k$ matrices with entries in $\mathbb F= \mathbb R,\mathbb C,\mathbb H$ for $\beta=1,2,4$. For large $M$ the right hand side of~\eqref{newman} can be evaluated using the law of large numbers.
Recently, it was pointed out in~\cite{Forrester:2015} that the method of~\cite{CN:1984} also can be used to extract the variances of the Lyapunov exponents. In this case, formula~\eqref{newman} is used to evaluate
\begin{equation}
\average[\big]{(\Lambda_k^\beta)^2}-\average[\big]{\Lambda_k^\beta}^2
=\sum_{n=1}^k\Var(\lambda_{N-n+1}^\beta)+2\sum_{1\leq i<j\leq k}\Cov(\lambda_{N-i+1}^\beta,\lambda_{N-j+1}^\beta).
\label{var}
\end{equation}
If the Lyapunov spectrum is non-degenerate then the covariances are expected to decay exponentially at large $M$, hence they may be neglected at leading order in $M$, see~\cite{Forrester:2015}. Note that if we are only interested in the largest Lyapunov exponent, i.e. $k=1$, then the second sum on the right hand side of~\eqref{var} disappears which was used much earlier in~\cite{CN:1984} to find the variance of the largest Lyapunov exponent. 

Also mixed products of Ginibre and inverse Ginibre matrices as well as products of truncated unitary matrices can be evaluated using either the probabilistic methods from~\cite{CN:1984} or the exactly solvable models from section~\ref{finiteNM} as in~\cite{ABK,Jesper2}. The mixed product case can be constructed from~\eqref{Lgauss} and will not be repeated here, see~\cite{Forrester:2015}.
We can consider unitary matrices truncated from $(N+\kappa)\times (N+\kappa)$ Haar distributed matrices to $N\times N$ sub-blocks (see the discussion in section~\ref{EVtrunc}) leading to~\cite{Forrester:2015}  
\begin{align}
\lambda_n^\beta&=\frac12\psi\bigg(\frac{\beta n}2\bigg)-\frac12\psi\bigg(\frac{\beta (\kappa+n)}2\bigg), \\
(\sigma_n^\beta)^2&=\frac14\psi'\bigg(\frac{\beta n}2\bigg)-\frac14\psi'\bigg(\frac{\beta (\kappa+n)}2\bigg).
\label{Ltrunc}
\end{align}
These quantities can also be found by direct evaluation of the jpdf from sections~\ref{SVtrunc} and~\ref{EVtrunc}. While the discussion given above involves solely square matrices, it is a straightforward task to include the parameters $\nu_j$, see~\cite{Jesper2}.

Although Lyapunov exponents are the standard choice for the characterisation of stability, a different possibility exists. Rather than investigating the singular values of the product matrix~\eqref{prod2}, we could look at the complex eigenvalues  $z_n$ ($n=1,\ldots,N$) of the product matrix~\eqref{prod2}. The absolute values of the eigenvalues are expected to grow exponentially for large $M$, although there is no equivalent of Oseledec's theorem in this case. It was therefore suggested in~\cite{ABK} to parametrise the eigenvalues as $z_n= e^{M\xi_n+i\theta_n}$, with $\xi_n$ and $\theta_n$ real. It was previously conjectured that the \emph{stability exponents} driving the exponential growth (decay) are identical to the Lyapunov exponents whenever the spectrum is non-degenerate~\cite{GSO:1987,ABK,Jesper2}. For the product of Ginibre matrices this was proved in~\cite{ABK,Jesper2} starting from eqs. 
\eqref{jpdf2}, \eqref{jpdf4} and \eqref{jpdf1real},
leading to \eqref{LyapunovPer} with identical means and variances \eqref{Lgauss} for the stability exponents
for $\beta=2,4$ and $1$, respectively. Note 
however that the corrections to this limit which were also computed in~\cite{ABK,Jesper2} differ from the corrections for the Lyapunov exponents
known for $\beta=2$ only.

It is natural to ask about the angular dependence of the complex eigenvalues for $M$ going to infinity as well. For $\beta=2$ the spectrum is trivially rotational invariant; for products of $\beta=4$ Ginibre matrices it was shown in~\cite{Jesper2} that the eigenvalues have a sine squared repulsion from the real axis, due to the pairing of complex conjugate eigenvalues. The most interesting case is undoubtedly $\beta=1$. For the product of real Ginibre matrices all eigenvalues becomes real when the number of matrices tends to infinity. This was first observed numerically in~\cite{arul} and for $N=2$, and later verified for general $N$ by explicit calculations in~\cite{PFreal}. It was found that for large $N$ the probability that all eigenvalues are real can be estimated by~\cite{PFreal}
\begin{equation}
\prob[\forall j:z_j\in\mathbb R]\sim\bigg(\frac{1}{\sqrt2}\bigg(\frac{\Gamma(1/M+1)\Gamma(1/2)}{\Gamma(1/M+1/2)}\bigg)^{M/4}\,\bigg)^{N^2},
\end{equation}
which tends to unity as $M$ goes to infinity. This was confirmed in~\cite{Jesper2}.

% ---------------------------------------

% ---------------------------------------

\section{Open Problems}
\label{concs}

\noindent
Several open problems can be easily identified. It would be desirable to allow for a wider choice of ensembles of random matrices that can be multiplied, while maintaining the exact solvability for finite products of $M$ factors at finite matrix size $\sim N$. Apart from allowing for classical Hermitian ensembles one generalisation relevant for 
universality is to include invariant non-Gaussian ensembles. As in the truncated unitary ensemble this immediately drops the independence amongst the matrix elements within each factor. On the other hand one could also drop the independence among different factors in the product. 
Furthermore, new limiting kernels may be accessible from the finite $N$ and $M$ results, such as by taking different double scaling limits.
 
Taking factors from the real Ginibre ensemble or studying truncated orthogonal matrices still presents a challenge. But even for the truncated unitary ensemble only a very limited parameter range of the exact solution at finite $N$ and $M$ has been explored in the weak non-unitarity limit. This and the orthogonal ensemble may provide a rich study ground, being potentially relevant in systems with absorption or with sources of decoherence.  Generally speaking the plethora of new mathematical results reviewed here has not yet been fully exploited in applications. Especially the new microscopic classes at the origin labelled by $M$ will probably require to go beyond standard areas such as combinatorics or telecommunications. The initial motivation to study toy models of dynamical systems in the beginning of the sixties may be very worthwhile revisiting under this new light of insights.\\

\noindent
{\bfseries Acknowledgements:} We would like to thank the organisers of this excellent workshop for creating a very stimulating atmosphere. We are indebted to all our coworkers for inspiring collaborations, which have lead to many of the results presented here. In particular we thank Peter Forrester, Mario Kieburg and Arno Kuijlaars for detailed comments on this manuscript.
This work was supported by SFB$|$TR12 (G.A.) and by IRTG 1132 (J.R.I.) of the German Science Foundation DFG.

\addcontentsline{toc}{section}{References}

\raggedright


\begin{thebibliography}{999}

\bibitem{FK60} H. Furstenberg and H. Kesten, Ann. Math. Stat. {\bfseries 31} (1960) 457.

\bibitem{Gin} J. Ginibre, 
%\textit{Statistical ensembles of complex, quaternion and real matrices.}  
J. Math. Phys. {\bfseries 6} (1965) 440.%--449.

\bibitem{JW:2004}
  R.A.~Janik, and W.~Wieczorek.
  %``Multiplying unitary random matrices--universality and spectral properties,''
  J. Phys. A {\bfseries 37} (2004) 6521 [arXiv:math-ph/0312043].

\bibitem{NN:2007}
  R.~Narayanan and H.~Neuberger,
  %``Universal properties of large N phase transitions in Wilson loops,''
  PoS LAT {\bfseries 2007} (2007) 272
  [arXiv:0709.4494].
  %%CITATION = ARXIV:0709.4494;%%

\bibitem{BN:2008}
  J.P.~Blaizot and M.A.~Nowak,
  %``Large N(c) confinement and turbulence,''
  Phys.\ Rev.\ Lett.\  {\bfseries 101} (2008) 102001
  [arXiv:0801.1859].
  %%CITATION = ARXIV:0801.1859;%%

\bibitem{CPV}  A. Crisanti, G. Paladin, and A. Vulpiani, \textit{Products of Random Matrices}, Springer, 1993.


\bibitem{Ralf} R.R. M\"uller, IEEE Trans. Inf. Theor. {\bfseries 48} (2002) 2086.

\bibitem{Karol} %Product of Ginibre matrices: Fuss-Catalan and Raney distributions
K. A. Penson, K. Zyczkowski,
Phys. Rev. E {\bfseries 83} (2011) 061118 [arXiv:1103.3453].

\bibitem{ABK} G. Akemann, Z. Burda, and M. Kieburg,
J. Phys. A: Math. Theor. {\bfseries 47} (2014) 395202 [arXiv:1406.0803]. 

\bibitem{Jesper2} J.R. Ipsen, arXiv:1412.3003.

\bibitem{GSO:1987}
  I. Goldhirsch, P.-L. Sulem, and S.A. Orszag. 
  %"Stability and Lyapunov stability of dynamical systems: A differential approach and a numerical method."
  Physica D {\bfseries  27} (1987) 311.

\bibitem{BJW} Z. Burda, R. A. Janik, and B. Waclaw, Phys. Rev. E {\bfseries 81} (2010) 041132 [arXiv:0912.3422].

\bibitem{GT} F. G\"otze and A. Tikhomirov, %On the Asymptotic Spectrum of Products of Independent Random Matrices, 
arXiv:1012.2710.

\bibitem{OS} S. O’Rourke and A. Soshnikov, Electr. J. Prob. {\bfseries 81} (2011) 2219 %-2245 
[arXiv:1012.4497].

\bibitem{Zrev} Z. Burda, 
%Free products of large random matrices - a short review of recent developments. 
J. Phys. Conf. Ser. {\bfseries 473} (2013) 012002 [arXiv:1309.2568].

\bibitem{Burda} Z. Burda, A. Jarosz, G. Livan, M.A. Nowak, and A. Swiech, Phys. Rev. E {\bfseries 82} (2010) 061114 [arXiv:1007.3594]; 
Acta Phys. Polon. B {\bfseries 42} (2011) 939 [arXiv:1103.3964].

\bibitem{ORSV} %Products of independent elliptic random matrices
S. O'Rourke, D. Renfrew, A. Soshnikov, and V. Vu, arXiv:1403.6080.

\bibitem{Andy} A.D. Jackson, B. Lautrup, P. Johansen, and M. Nielsen, Phys. Rev. E {\bfseries 66} (2002) 066124 [arXiv:physics/0202037].

\bibitem{ab} G. Akemann and Z. Burda, J. Phys. A: Math. Theor. {\bfseries 45}, 465201 (2012) [arXiv:1208.0187].

\bibitem{as} G. Akemann and E. Strahov, J. Stat. Phys. {\bfseries 151} (2013) 987 [arXiv:1211.1576]. 

\bibitem{Jesper} J.R. Ipsen, J. Phys. A {\bfseries 46} (2013) 265201 [arXiv:1301.3343].

\bibitem{akw} G. Akemann, M. Kieburg, and L. Wei, J. Phys. A {\bfseries 46} (2013) 275205 [arXiv:1303.5694].

\bibitem{aik} G. Akemann, J. Ipsen, and M. Kieburg, Phys. Rev. E {\bfseries 88} (2013) 052118 [arXiv:1307.7560].

\bibitem{Springer} M.D. Springer and W.E. Thompson, %{\itshape The distribution of products of beta, gamma and Gaussian random variables.} 
SIAM J. Appl. Math. {\bfseries 18} (1970) 721.

\bibitem{Fischmann}
J. Fischmann, W. Bruzda, B.A. Khoruzhenko, H.-J. Sommers, and K. Zyczkowski, 
%\textit{Induced Ginibre ensemble of random matrices and quantum operations.} 
J. Phys. A {\bfseries 45} (2012) 075203 [arXiv:1107.5019].

\bibitem{ARRS} K. Adhikari, N.K. Reddy, T.R. Reddy, and K. Saha,  arXiv:1308.6817.

\bibitem{IK} J. R. Ipsen and M. Kieburg, Phys. Rev. E {\bfseries 89} (2014) 032106 [arXiv:1310.4154].

\bibitem{GL:1996}
  G.H.~Golub and C.F. van Loan,
  \textit{Matrix computations}, 3rd Edition, 
  JHU Press, 1996.

\bibitem{es} E. Strahov, unpublished notes (2013).

\bibitem{ais} G. Akemann, J.R. Ipsen, and E. Strahov, Random Matrices: Th. Appl. {\bfseries 3}
%Vol. 3, No. 4 
(2014) 1450014 	[arXiv:1404.4583].

\bibitem{PFreal} P.J. Forrester, J. Phys. A {\bfseries 47} (2014) 065202 [arXiv:1309.7736]. 

\bibitem{ABKN} G. Akemann, Z. Burda, M. Kieburg, and T. Nagao, 
J. Phys. A: Math. Theor. {\bfseries 47} (2014) 255202
[arXiv:1310.6395].

\bibitem{LW} %Universality for products of random matrices I: Ginibre and truncated unitary cases
%Dang-Zheng Liu, Yanhui Wang
D.-Z. Liu and Y. Wang, 	arXiv:1411.2787.

\bibitem{KS} K. Zyczkowski and H.-J. Sommers, J. Phys. {\bfseries A33} (2000) 2045 [arXiv:chao-dyn/9910032].

\bibitem{KS2} B.A. Khoruzhenko, H.-J. Sommers, and K. Zyczkowski, Phys. Rev. {\bfseries E82} (2010) 040106(R) [arXiv:1008.2075].

\bibitem{Hough}
 J.B.~Hough; M.~Krishnapur, Y.~Peres; B.~Vir$\acute{\mbox{a}}$g. \textit{Zeros of Gaussian analytic functions and determinantal point processes}, 
%University Lecture Series, 51. American Mathematical Society, 
AMS, Providence, RI, 2009.

\bibitem{Rider}
B. Rider,  
%\textit{Deviations from the circular law.} 
Probab. Theory Related Fields {\bfseries 130} (2004) 337.

\bibitem{JQ} %Spectral Radii of Large Non-Hermitian Random Matrices
T. Jiang and Y. Qi, arXiv:1411.1833.

\bibitem{lun} L. Zhang,
%A note on the limiting mean distribution of singular values for products of two Wishart random matrices,
J. Math. Phys. {\bfseries 54} (2013) 083303
[arXiv:1305.0726].

\bibitem{AZ}
A.B.J. Kuijlaars and L. Zhang,
%\textit{Singular values of products of Ginibre random matrices, multiple orthogonal polynomials and hard edge scaling limits.}
Commun. Math. Phys. {\bfseries 332} (2014) 759, %-781
[arXiv:1308.1003].

\bibitem{MB1} M. Bertola, M. Gekhtman, and J. Szmigielski, %The Cauchy two-matrix model, 
Commun. Math. Phys. {\bfseries 287} (2009) 983; %–1014.
arXiv:1211.5369.

\bibitem{MB2} 
%Universality conjecture and results for a model of several coupled positive-definite matrices
M. Bertola and T. Bothner, arXiv:1407.2597.

\bibitem{FK:2014}
  P.J. Forrester and M. Kieburg,
  %``Relating the Bures measure to the Cauchy two-matrix model,''
%  Preprint (2014)
arXiv:1410.6883.

\bibitem{ArnoDries} %Singular values of products of random matrices and polynomial ensembles
A.B.J. Kuijlaars and D. Stivigny, 	
Random Matrices: Th. Appl. {\bfseries 03} (2014) 1450011
[arXiv:1404.5802].

\bibitem{Muttalib:1995}
  K.~A.~Muttalib,
  %``Random matrix models with additional interactions,''
  J. Phys. A {\bfseries 28} (1995) L159.

\bibitem{Borodin} A. Borodin,
Nucl. Phys. B {\bfseries 536} (1998) 704 %–732
[arXiv:math/9804027].

\bibitem{Kuijlaars:2015}
  A.B.J. Kuijlaars,
  %``Transformations of polynomial ensembles.''
%  Preprint (2015)
arXiv:1501.05506.

\bibitem{PF14}  
%Eigenvalue statistics for product complex Wishart matrices
P.J. Forrester, J. Phys. A: Math. Theor. {\bfseries 47} (2014) 345202 	[arXiv:1401.2572].

\bibitem{TD}	
%Asymptotics for characteristic polynomials of Wishart type products of complex Gaussian and truncated unitary random matrices
T. Neuschel and D. Stivigny, arXiv:1407.2755.

\bibitem{ES14}
% Differential equations for singular values of products of Ginibre random matrices
E. Strahov, J. Phys. A: Math. Theor {\bfseries 47} (2014) 325203 [1403.6368].

\bibitem{lundang} %Bulk and soft-edge universality for singular values of products of Ginibre random matrices
%Dang-Zheng Liu, Dong Wang, Lun Zhang,
D.-Z. Liu, D. Wang, and L. Zhang, arXiv:1412.6777.

\bibitem{KKS}
  M. Kieburg, A.B.J. Kuijlaars, and D. Stivigny,
  %``Singular value statistics of matrix products with truncated unitary matrices,''
%  Preprint (2015)
arXiv:1501.03910.

 \bibitem{Mario15} M. Kieburg,
arXv:1502.00550

\bibitem{PF13} P.J. Forrester, J. Stat. Phys. {\bfseries 151} 796 (2013) [arXiv:1206.2001].

\bibitem{Kargin}
  V. Kargin,
  %"On the largest Lyapunov exponent for products of Gaussian matrices." 
  J. Stat. Phys. {\bfseries 157} (2014) 70 [arXiv:1306.6576]. 

\bibitem{arul} A. Lakshminarayan, J. Phys. A {\bfseries 46} (2013) 152003  [arXiv:1301.7601].

\bibitem{James}
J.C. Osborn, Phys. Rev. Lett. {\bfseries 93} (2004) 222001 [arXiv:hep-th/0403131].

\bibitem{a05}
G. Akemann,
  %``The complex Laguerre symplectic ensemble of non-Hermitian matrices,''
  Nucl. Phys. B {\bfseries 730} (2005) 253 [arXiv:hep-th/0507156].

\bibitem{aps}
G. Akemann, M.J. Phillips, and H.J. Sommers,
  %``The chiral Gaussian two-matrix ensemble of real asymmetric matrices,''
  J. Phys. A {\bfseries  43} (2010) 085211 [arXiv:0911.1276]. % [hep-th]].

\bibitem{A05acta} G. Akemann,
%Non-Hermitian extensions of Wishart random matrix ensembles
Acta Phys. Polon. {\bf B 42} (2011)  0901
[arXiv:1104.5203].

\bibitem{as2} G. Akemann and E. Strahov, in preparation.

\bibitem{Mehta} M.L. Mehta, {\itshape Random Matrices}, 3rd Edition, Elsevier, 2004.

\bibitem{Bateman:1953}
  Bateman Manuscript Project, Arthur Erdélyi (ed.),
  \textit{Higher Transcendental Functions}, Vol. 1,
  McGraw-Hill, 1953.

\bibitem{Kostlan}
E. Kostlan, 
% \textit{On the spectra of Gaussian matrices.} Directions in matrix theory (Auburn, AL, 1990). 
Linear Algebra Appl. {\bfseries 162/164} (1992), 385.%-388.

\bibitem{apshi}
G. Akemann, M.J. Phillips, and L. Shifrin, 
%\textit{Gap probabilities in non-Hermitian random matrix theory.} 
J. Math. Phys. {\bfseries 50} (2009)%, no. 6, 
063504 [arXiv:0901.0897].

\bibitem{Kanzieper:2002}
  E.~Kanzieper,
  %``Eigenvalue correlations in Ginibre's nonHermitean random matrices at beta = 4,''
  J.\ Phys.\ A {\bfseries 35} (2002) 6631
  [arXiv:cond-mat/0109287].
  %%CITATION = COND-MAT/0109287;%%

\bibitem{Krishnapur}
M. Krishnapur, %Zeros of random analytic functions, 
Ann. Prob. {\bfseries 37} (2009), 314%.{346.

\bibitem{Anthony}
A. Mays, 
%A real quaternion spherical ensemble of random matrices
J. Stat. Phys.
%October 2013, Volume 153, Issue 1, pp 48-69 
{\bfseries 153} (2013) 48
[arXiv:1209.0888].

\bibitem{FM}
  %    Pfaffian point process for the Gaussian real generalised eigenvalue problem
  P.J. Forrester and A. Mays, 
  Probab. Theory and Related Fields
  %October 2012, Volume 
  {\bfseries 154} (2012) 1 
 [arXiv:0910.2531].

\bibitem{IZ:1980}
  C.~Itzykson and J.B.~Zuber,
  %``The Planar Approximation. 2.,''
  J.\ Math.\ Phys.\  {\bf 21} (1980) 411.
  %%CITATION = JMAPA,21,411;%%

\bibitem{Ismail:2005}
  M.E.H. Ismail,
  \textit{Classical and Quantum Orthogonal Polynomials in One Variable},
  Cambridge University Press, 2005.

\bibitem{Kuijlaars:2010}
  A.B.J. Kuijlaars,
  %``Multiple orthogonal polynomial ensembles,''
  Contemp. Math. {\bfseries 507} (2010) 155.

\bibitem{GKT:2014}
  F. G\"otze, H. K\"osters, and A. Tikhomirov.
  %``Asymptotic spectra of matrix-valued functions of independent random matrices and free probability,''
%  Preprint (2014) [
arXiv:1408.1732.

\bibitem{AGT:2010}
  N. Alexeev, F. G\"otze, and A. Tikhomirov.
  %``Asymptotic distribution of singular values of powers of random matrices.'' 
  Lith. Math. J. {\bfseries  50} (2010) 121.

\bibitem{BBCC:2011}
  T. Banica, S.T. Belinschi, M. Capitaine, and B. Collins.
  %``Free Bessel laws.''
  Canad. J. Math.{\bfseries  63} (2011) 3.

\bibitem{NS:2006}
  A. Nica and R. Speicher.
  \textit{Lectures on the combinatorics of free probability}, 
  Cambridge University Press, Cambridge, 2006.

\bibitem{Forrester:2010}
  P.J. Forrester, \textit{Log-gases and random matrices},
  Princeton University Press, 2010.

\bibitem{AHM:2011}
  Y. Ameur, H. Hedenmalm and N. Makarov, 
  Duke Math. J. {\bfseries  159} (2011) 31 [arXiv:0807.0375].

\bibitem{Berman:2008}
  R.J. Berman,
  Comm. Math. Phys. {\bfseries  327} (2014) 1 [arXiv:0811.3341].

\bibitem{TV:2012}
  T. Tao, and V. Vu. 
  %"Random matrices: Universality of local spectral statistics of non-Hermitian matrices."
%  Preprint (2012) [
arXiv:1206.1893.

\bibitem{PP:2014}
  R. Prakash, and A. Pandey.
  %"Universal spectral correlations in ensembles of random normal matrices." 
  arXiv:1412.6642.

\bibitem{Rider03}
B. Rider,
% A limit Theorem at the edge of a non-Hermitian random matrix ensemble,
J. Phys. A {\bfseries 36} (2003) 3401. %-3409

\bibitem{BGS:2014}
  M. Bertola, M. Gekhtman, and J. Szmigielski, 
  %``Cauchy-Laguerre two-matrix model and the Meijer-G random point field,''
  Comm. Math. Phys. 326 (2014) 111
  [arXiv:1211.5369].

 \bibitem{ADMN:1997}
   G.~Akemann, P.H.~Damgaard, U.~Magnea and S.~Nishigaki,
   %``Universality of random matrices in the microscopic limit and the Dirac operator spectrum,''
   Nucl.\ Phys.\ B {\bf 487} (1997) 721
   [hep-th/9609174].
   %%CITATION = HEP-TH/9609174;%%
 
 \bibitem{KF:1998}
   E. Kanzieper and V. Freilikher.
   %"Random-matrix models with the logarithmic-singular level confinement: method of fictitious fermions,"
   Philos. Magazine B {\bfseries 77} (1998) 1161 [arXiv:cond-mat/9704149].
 
 \bibitem{KV:2002}
   A.B.J. Kuijlaars and M. Vanlessen.
   %"Universality for eigenvalue correlations from the modified Jacobi unitary ensemble."
   Int. Math. Res. Notices {\bfseries 30} (2002) 1575 [arXiv:math-ph/0204006].
 
 \bibitem{KV:2003}
   A.B.J. Kuijlaars and M. Vanlessen.
   %"Universality for eigenvalue correlations at the origin of the spectrum."
   Comm. Math. Phys. {\bfseries  243} (2003) 163 [arXiv:math-ph/0305044].
 
 \bibitem{Lubinsky:2008}
   D.S. Lubinsky,
   %``Universality limits at the hard edge of the spectrum for measures with compact support,''
   Int. Math. Res. Notices (2008) 39.

\bibitem{FL} 
  %Raney distributions and random matrix theory
  P.J. Forrester, D.-Z. Liu,
  J. Stat. Phys. (2014) 1 [arXiv:1404.5759].

\bibitem{LSW:2011}
  D.-Z. Liu, C. Song, and Z.-D. Wang.
  %``On explicit probability densities associated with Fuss-Catalan numbers.''
  Proc. Amer. Math. Soc. {\bfseries 139} (2011) 3735 [arXiv:1008.0271 [math.PR]].

\bibitem{Biane:1998}
  P. Biane. 
  %``Processes with free increments.''
  Math. Z. {\bfseries  227} (1999) 143.

\bibitem{HM:2013}
  U. Haagerup and S. M\"oller. 
  %``The law of large numbers for the free multiplicative convolution.''
  In \textit{Operator algebra and dynamics}, Springer, 2013.

\bibitem{TN} T. Neuschel, 
%Plancherel-Rotach formulae for average characteristic polynomials of products of Ginibre random matrices and the Fuss-Catalan distribution, 
Random Matrices: Th. Appl. {\bfseries 3} (2014) 1450003 [arXiv:1311.0365].

\bibitem{Oseledec:1968}
  V.I. Oseledec,
  %``A multiplicative ergodic theorem. Lyapunov characteristic numbers for dynamical systems,'' 
  Trans. Moscow Math. Soc. {\bfseries  19} (1968) 197.

\bibitem{Raghunathan:1979}
  M.S. Raghunathan,
  %``A proof of Oseledec’s multiplicative ergodic theorem,'' 
  Isreal J. Math. {\bfseries  32} (1979) 356.


\bibitem{CKN:1986}
  J.E. Cohen, H. Kesten, and Ch.M. Newman (eds.) 
  \textit{Random Matrices and Their Applications},
  AMS, 1986.

\bibitem{Virster:1970}
  A.D. Virtser,
   %"Central limit theorem for semisimple Lie groups."
  Theory Prob. App. {\bfseries  15} (1970) 667.


\bibitem{CN:1984}
  J.E. Cohen, and Ch.M. Newman,
  %"The stability of large random matrices and their products."
  Ann. Prob. {\bfseries 12} (1984) 283.



\bibitem{Newman:1986}
  Ch.M. Newman,
  %"The distribution of Lyapunov exponents: Exact results for random matrices." 
  Comm. Math. Phys. {\bfseries  103} (1986) 121.

\bibitem{Forrester:2015}
  P.J. Forrester,
%  Preprint (2015) [
arXiv:1501.05702.










\end{thebibliography}
\end{document}